\documentclass[journal]{IEEEtran}
\label{package}
\usepackage{graphics}
\usepackage{graphicx}
\usepackage{subfigure}
\usepackage{multirow}
\usepackage{array}
\usepackage{soul, xcolor}
\usepackage[ruled,vlined,linesnumbered]{algorithm2e}
\usepackage{changes}
\usepackage{MnSymbol}
\usepackage{amsthm} 
\usepackage{url}
\usepackage{caption} 
\usepackage[hidelinks]{hyperref}
\usepackage{booktabs}
\usepackage[utf8]{inputenc}
\usepackage{csquotes}
\usepackage[T1]{fontenc}

\definecolor{bio-color}{RGB}{230,242,255}
\definecolor{comp-color}{RGB}{255,230,230}
\definecolor{phy-color}{RGB}{230,255,230}
\definecolor{chem-color}{RGB}{255,230,255}

\makeatletter
\newcommand{\linelabel}[1]{%
  \refstepcounter{AlgoLine}% 将行号计数器前进一次并使其可被 \label 捕获
  \label{#1}% 打标签（指向即将出现的行号）
  \addtocounter{AlgoLine}{-1}% 把计数器退回，保证 algorithm2e 在真正输出该行时使用相同的行号
}
\makeatother

\newtheorem{example}{Example}[section]
\newtheorem{property}{Property}[section]
\newtheorem{theorem}{Theorem}[section]
\newtheorem{lemma}{Lemma}[section]
\newtheorem{corollary}[theorem]{Corollary} % 定义推论环境

\newtheorem{definition}{Definition}[section]
\newtheorem*{problem}{Problem statement} 
\newcounter{nestingLevel}

\newcommand{\Rmnum}[1]{\expandafter \romannumeral #1@}

\begin{document}
%
% paper title
% Titles are generally capitalized except for words such as a, an, and, as,
% at, but, by, for, in, nor, of, on, or, the, to, and up, which are usually
% not capitalized unless they are the first or last word of the title.
% Linebreaks \\ can be used within to get better formatting as desired.
% Do not put math or special symbols in the title.
\title{
EAIFD: A Fast and Scalable Algorithm for Incremental Functional Dependency Discovery
}
%
%
% author names and IEEE memberships
% note positions of commas and nonbreaking spaces ( ~ ) LaTeX will not break
% a structure at a ~ so this keeps an author's name from being broken across
% two lines.
% use \thanks{} to gain access to the first footnote area
% a separate \thanks must be used for each paragraph as LaTeX2e's \thanks
% was not built to handle multiple paragraphs
%

\author{Yajuan~Xu,
	Xixian~Han,
	Xiaolong~Wan
	% <-this % stops a space
\IEEEcompsocitemizethanks{
	\IEEEcompsocthanksitem Yajuan~Xu, Xixian~Han and Xiaolong~Wan are with the School of Computer Science and Technology, Harbin Institute of Technology, China. (e-mail: xuyajuan@stu.hit.edu.cn, hanxx@hit.edu.cn, wxl@hit.edu.cn)}% <-this % stops a space
\thanks{Manuscript received XX XX, XXXX; revised XX XX, XXXX.}}

% The paper headers
%\markboth{IEEE Transactions on Knowledge and Data Engineering,~Vol.~XX, No.~XX, XX~XXXX}%
%{Xu \MakeLowercase{\textit{et al.}}: A Fast and Scalable Algorithm for Incremental Functional Dependency Discovery}

\maketitle

% As a general rule, do not put math, special symbols, or citations
% in the abstract or keywords.
\begin{abstract}
Functional dependencies (FDs) are fundamental integrity constraints in relational databases, but discovering them under incremental updates remains challenging. While static algorithms are inefficient due to full re-execution, incremental algorithms suffer from severe performance and memory bottlenecks. To address these challenges, this paper proposes EAIFD, a novel algorithm for incremental FD discovery. EAIFD maintains the partial hypergraph of difference sets and reframes the incremental FD discovery problem into minimal hitting set enumeration on hypergraph, avoiding full re-runs. EAIFD introduces two key innovations. First, a multi-attribute hash table ($MHT$) is devised for high-frequency key-value mappings of valid FDs, whose memory consumption is proven to be independent of the dataset size. Second, two-step validation strategy is developed to efficiently validate the enumerated candidates, which leverages $MHT$ to effectively reduce the validation space and then selectively loads data blocks for batch validation of remaining candidates, effectively avoiding repeated I/O operations. Experimental results on real-world datasets demonstrate the significant advantages of EAIFD. Compared to existing algorithms, EAIFD achieves up to an order-of-magnitude speedup in runtime while reducing memory usage by over two orders-of-magnitude, establishing it as a highly efficient and scalable solution for incremental FD discovery.
\end{abstract}

% Note that keywords are not normally used for peerreview papers.

\IEEEpeerreviewmaketitle

\section{Introduction}

\IEEEPARstart{F}{unctional} dependencies~(FDs) are fundamental integrity constraints in relational databases~\cite{DBLP:books/mg/SKS20}, which are widely applied in many areas, e.g., schema design and normalization~\cite{DBLP:conf/edbt/PapenbrockN17,DBLP:journals/tods/WeiL21}, query optimization~\cite{DBLP:journals/vldb/KossmannPN22}, data cleaning~\cite{DBLP:conf/icde/BohannonFGJK07} and data integration~\cite{DBLP:books/daglib/0029346}. Formally, for a relation instance \textit{r} of schema \textit{R}, a functional dependency (FD) $X \rightarrow A$ is valid iff for any two tuples in $r$, if their values on $X$ are the same, then they must have the same value on $A$, where $X \subseteq R$ and $A \in R$ are referred to as left-hand side (LHS) and right-hand side (RHS), respectively. The FD $X \rightarrow A$ is non-trivial if $A \notin X$, and minimal if there is no proper subset $X'\subset X$ such that $X' \rightarrow A$ holds in $r$. FD discovery focuses on \textit{the complete set $\mathcal{F}$ of minimal and non-trivial FDs}.

\begin{figure}[!htbp]              
	\centering
        \includegraphics[scale = 0.34]{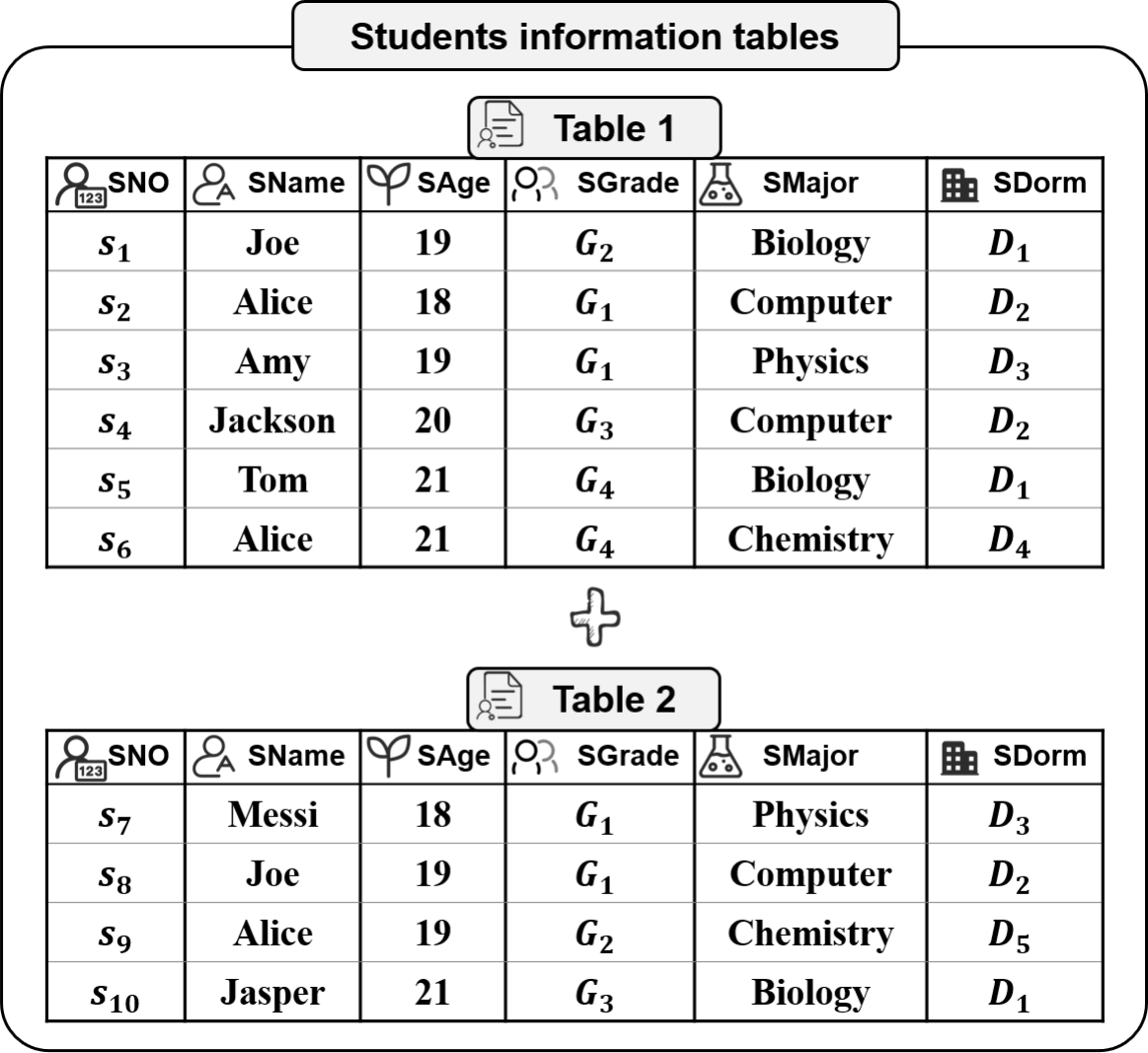}
        \caption{\small The figure of two student information tables. Table $1$ contains $6$ student information tuples, Table $2$ contains $4$ student information tuples, and the tuples in Table $2$ are incrementally added to Table $1$.}
        \label{fig:students}
\end{figure}

\begin{example}
	\label{example:introduction_exp}
	A student table is shown in Table 1 of Fig.~\ref{fig:students}. $\text{SName} \to \text{SAge}$ is invalid, as the tuple pairs ($s_2$ and $s_6$) share the same name but have different ages. $\{\text{SNO}, \text{SName}\} \to \text{SMajor}$ is a valid but not minimal FD. And $\mathcal{F} = \{  \text{SNO} \to \text{SName},\ \text{SNO} \to \text{SAge},\   \text{SNO} \to \text{SGrade},\  \text{SNO} \to \text{SMajor},\  \text{SNO} \to \text{SDorm},\   \text{SMajor} \to \text{SDorm} \}$. 
\end{example}

The real-life datasets typically grow continuously through the addition of tuples. For example, the modern newspaper database\footnote{\url{https://www.nlcpress.com/DigitalPublishingView.aspx?DPId=10}.} adds at least $100$ new newspaper titles annually. The Ncvoter dataset\footnote{\url{https://www.ncsbe.gov/results-data/voter-history-data}.} is incrementally updated as North Carolina counties finalize voter histories after each election. The incremental updates of a database often lead to changes in $\mathcal{F}$. As illustrated in Fig.~\ref{fig:students}, when new tuples from Table $2$ are added to Table $1$, previously valid FD $\text{SMajor} \to \text{SDorm}$ is invalid, due to the tuple pair $(s_6,s_9)$. $\{\text{SGrade}, \text{SMajor}\} \to \text{SDorm}$ becomes a new minimal valid FD. 

Traditional FD discovery algorithms \cite{DBLP:journals/cj/HuhtalaKPT99,DBLP:journals/is/NovelliC01,DBLP:conf/cikm/AbedjanSN14,DBLP:conf/sigmod/PapenbrockN16,DBLP:journals/pacmmod/BleifussPBSN24,DBLP:journals/dke/LiuYLW13,DBLP:journals/tkde/WanHWL24,DBLP:conf/dawak/WyssGR01,DBLP:journals/datamine/YaoH08,DBLP:conf/icde/WeiL19} are designed for static datasets and must be re-executed upon data inserts, rendering them impractical for continuous data growth. Therefore, researchers have explored FD discovery on incremental databases. While incremental algorithms, DynFD~\cite{DBLP:conf/edbt/SchirmerP0NHMN19} and DHSFD~\cite{DBLP:conf/icde/XiaoYTMW22}, perform better than static FD discovery algorithms, they still face the following challenges. Specifically, (1) \textbf{Long preprocessing time.} Both DynFD and DHSFD require costly preprocessing, with time complexity of $O(m \cdot 2^{m})$ and $O(n^2)$, which is impractical in real-life scenarios; (2) \textbf{Poor scalability on wide schema.} As $|R|$ increases, the candidate search space grows exponentially, degrading validation and maintenance cost; (3) \textbf{Excessive memory consumption.} They need to maintain many auxiliary data structures, which grow rapidly with data and cause high memory usage.

To address these challenges, this paper presents EAIFD, an \underline{e}fficient \underline{a}lgorithm for \underline{i}ncremental \underline{FD}s discovery, that provides fast initialization, scalability on high-dimensional datasets, and low memory consumption. EAIFD systematically tackles the aforementioned challenges through \textit{three key contributions}. \textbf{Firstly}, EAIFD addresses the challenge of long preprocessing time by performing a sampling-based initialization. It computes difference sets only on the sampled subset and builds partial hypergraphs, which significantly reduces $O(n^2)$ time complexity and greatly reduces preprocessing time. EAIFD generates candidate FDs by enumerating hitting sets on partial hypergraphs. \textbf{Secondly}, EAIFD optimizes the validation phase to reduce the high costs caused by the exponential growth of candidates. Specifically, it employs iterative grouping hash validation (IGHV) combined with a multi-attribute hash table ($MHT$) that stores mappings between LHS and RHS values. After data updates, IGHV compares the updated data with items in pre-built $MHT$ and restricts validation to relevant blocks and validates candidates in batches. Valid FDs are added to $\mathcal{F}$, while invalid FDs are used to update the hypergraph and iteratively generate new candidates. \textbf{Thirdly}, to control memory usage, EAIFD applies a \textit{high-frequency mapping items preservation strategy}, keeping only mapping items that appear above a specified frequency threshold, ensuring $MHT$ remains lightweight and balancing memory consumption with performance efficiency. Comprehensive experiments are conducted on real-life datasets. The experimental results verify that EAIFD is more efficient than the incremental FD discovery algorithms. Notably, compared with the existing algorithms, EAIFD achieves up to an order-of-magnitude speedup in runtime, while reducing memory usage by over two orders of magnitude.

The rest of this paper is organized as follows. Section~\ref{sec:relatedWork} surveys the related work, followed by problem statement and preliminaries in Section~\ref{sec:problemDefinition}. Section~\ref{sec:algorithm} introduces EAIFD algorithm. Section~\ref{sec:performanceEvaluation} provides a comprehensive experimental evaluation. Section~\ref{sec:conclusion} concludes the paper.

\section{Related Work}\label{sec:relatedWork} 
This section surveys the related work of FD discovery, and introduces two related research lines to our work: (\romannumeral1) FD discovery algorithms designed for static datasets, (\romannumeral2) FD discovery algorithms that support incremental datasets update.

\subsection{FD discovery for static data}\label{subsec:staticFDDiscover}

Researchers propose many FD discovery algorithms for static relational databases~\cite{DBLP:journals/cj/HuhtalaKPT99,DBLP:journals/is/NovelliC01,DBLP:journals/aicom/FlachS99,DBLP:conf/edbt/LopesPL00,DBLP:journals/dke/LiuYLW13,DBLP:conf/cikm/AbedjanSN14,DBLP:conf/sigmod/PapenbrockN16,DBLP:journals/pacmmod/BleifussPBSN24}, which consist of two key phases: (\romannumeral1) candidate FDs generation and (\romannumeral2) candidate FDs validation~\cite{DBLP:journals/tkde/WanHWL24}. The existing algorithms improve FD discovery efficiency from different aspects and can be roughly divided into the following three categories.

\textbf{Attribute-based algorithms}~\cite{DBLP:journals/cj/HuhtalaKPT99,DBLP:journals/pvldb/PapenbrockEMNRZ15,DBLP:conf/cikm/AbedjanSN14,DBLP:journals/dke/LiuYLW13,DBLP:journals/datamine/YaoH08,DBLP:journals/is/NovelliC01} enumerate candidate FDs in powerset lattices of attribute combinations and traverse lattices with level-wise or depth-first strategies. They usually use position list indexes (PLIs) to validate candidate FDs with many pruning rules to reduce the search space. They perform well on datasets with many tuples, but due to their candidate-driven search strategy, they scale poorly on datasets with many attributes.

\textbf{Tuple-based algorithms}~\cite{DBLP:journals/pacmmod/BleifussPBSN24,DBLP:conf/dawak/WyssGR01,DBLP:journals/aicom/FlachS99,DBLP:conf/edbt/LopesPL00} are based on the value comparisons of all tuple pairs. These algorithms discover FDs by constructing agree sets of attributes with identical tuple values and difference sets of attributes with distinct values. These algorithms scale well with the number of attributes. However, comparing all tuple pairs often leads to a prohibitive $O(n^2)$ complexity, which severely degrades performance on long datasets.

\textbf{Hybrid algorithms}~\cite{DBLP:conf/sigmod/PapenbrockN16,DBLP:conf/icde/WeiL19} integrate the advantages of both attribute-based and tuple-based algorithms. By employing a row-based strategy on sampled data and a column-based strategy on the full dataset, they effectively avoid numerous ineffective candidate FDs validations and tuple comparisons. Experimental results demonstrate that this hybrid design outperforms both attribute-based and tuple-based algorithms.

\subsection{FD discovery for incremental data}\label{subsec:incrementalFDDiscovery}

Traditional static FD discovery algorithms must re-execute the entire discovery process after each data update. In recent years, different researchers have turned their efforts to developing FD discovery algorithms suitable for incremental data scenarios. These methods aim to discover the complete set of minimal and non-trivial FDs at a lower computational cost after data updates without needing to re-execute the entire algorithm. DynFD~\cite{DBLP:conf/edbt/SchirmerP0NHMN19} and DHSFD~\cite{DBLP:conf/icde/XiaoYTMW22} are proposed for incremental FD discovery, which perform incremental FD discovery from the attribute and tuple perspectives, respectively.

\textbf{Attribute-based algorithm:} DynFD is the first algorithm capable of maintaining the complete set of minimal non-trivial FDs for incrementally updated data. Instead of frequently recomputing all FDs, DynFD detects changes in the data and infers their impact on FDs by incrementally updating the FDs set based on prior results and a batch of update operations. It maintains a positive cover of minimal FDs and a negative cover of maximal non-FDs. After incremental data updates, DynFD validates whether the most general FDs still hold. If an FD is invalid, it is removed from the positive cover, added to the negative cover, and specialized by adding new attributes to its LHS. The specialized candidates are then validated using only PLI clusters that include newly inserted tuples, improving efficiency. If the specialized FD holds, it is added to the positive cover. When the proportion of invalidated FDs in a validation round exceeds $10\%$, DynFD switches to a violation-driven strategy that compares new tuples with relevant existing records to locate invalid FDs. For each invalid FD, the positive and negative covers are updated accordingly to maintain consistency.

\textbf{Tuple-based algorithm:} DHSFD transforms the problem of incremental FD discovery into a dynamic hitting set enumeration problem over hypergraphs constructed from difference sets of entire dataset. DHSFD maintains auxiliary structures, including PLIs and difference sets with weights. After incremental data updates, PLIs are used to update the difference sets and their weights. These updates are treated as edge additions in the hypergraph, where each edge represents a difference set. DHSFD then computes minimal hitting sets over the updated hypergraph. It validates minimality using critical edges and extends candidate FDs using a depth-first strategy called WalkDown. The resulting minimal hitting sets form the LHSs of newly discovered minimal FDs. By comparing only tuples in clusters containing incremental tuples and avoiding reconstruction of the search tree, DHSFD achieves higher efficiency than static algorithms.

\textbf{Discussion.} This paper focuses on the problem of FD discovery for incremental data, which are continuously updated by adding tuples. Based on the current data scenarios, current incremental FD discovery faces the following challenges.

\textit{Preprocessing time is too long.} In the preprocessing phase, DynFD derives the negative cover from the positive cover, while DHSFD needs to compare all the tuples pair to compute difference sets of the dataset, this step has a time complexity $O(n^2)$ for a dataset with $n$ tuples. When the dataset or $|\mathcal{F}|$ is large, both methods severely impact the startup time.  

\textit{Inefficient algorithm performance.} DynFD is inefficient in high-dimensional datasets due to the exponential growth in the search space and increased memory usage for maintaining positive and negative covers. When the invalidation rate exceeds a threshold, DynFD switches to violation-driven validation and the cost of traversing PLI clusters containing new tuples becomes extremely expensive. DHSFD still faces a significant increase in the cost of updating its difference sets and hypergraphs.

\textit{Auxiliary data structure memory space consumption is high.} Two algorithms often rely on auxiliary data structures that consume a large amount of memory. DynFD maintains the positive cover, negative cover, position list indexes (PLIs), and dictionary encodings. Likewise, DHSFD needs to maintain PLIs, difference sets and hypergraphs. As new data is continuously appended, these structures grow in size, resulting in a significant increase in memory usage.

Motivated by these limitations, our goal is to design an incremental FD discovery algorithm that supports high-dimensional datasets, avoids expensive preprocessing, and operates with minimal memory overhead.

\section{Problem definition}
\label{sec:problemDefinition}

Let \textit{r} be a relation instance over schema \textit{R}. We denote incremental data as $\Delta r$. After incremental update, the updated dataset is denoted as $r \cup \Delta r$. $\forall t \in r$, we use $t[X] $ or $t[A] $ to denote the projection of $t$ on attribute set $X \subseteq R$ or single attribute $A \in R$, respectively.

\begin{definition}
	\label{definition 1}
	\textbf{(FD).} The FD $X \rightarrow A$ with $X \subseteq R$ and $A \in R$ is valid for instance $r$ of $R$, iff $\forall t_{1}, t_{2}\in r: t_{1}[X] =t_{2}[X] \Rightarrow t_{1}[A] = t_{2}[A]$. If there is no $Y \subset X,Y \rightarrow A$ and $A \notin X$, then $X \rightarrow A$ is minimal and non-trivial.
\end{definition}

Given FDs $X \rightarrow A$ and $Y \rightarrow A$ where $X \subset Y$, $X \rightarrow A $ is a \textbf{generalization} of $Y \rightarrow A $ and $Y \rightarrow A $ is a \textbf{specialization} of $X \rightarrow A $. FD discovery focuses on the complete set $\mathcal{F}$ of minimal and non-trivial FDs, as all other valid FDs can be derived from them using Armstrong's axioms~\cite{DBLP:books/daglib/0020812}.

\begin{definition}
	\label{definition 2}
	\textbf{(Violated tuple pair).} If a candidate FD $X \rightarrow A$ is invalid, then there must exist a pair of tuples $t_1, t_2 \in r$ where $t_1[X] = t_2[X]$ but $t_1[A] \ne t_2[A]$. The $\left ( t_1, t_2\right ) $ is called a violated tuple pair for the candidate FD $X \rightarrow A$.
\end{definition}

This paper transforms FD discovery into hitting set enumeration in hypergraphs, following tuple-based algorithms ~\cite{DBLP:conf/dawak/WyssGR01,DBLP:journals/dke/MannilaR94,DBLP:conf/icde/XiaoYTMW22}. The necessary definitions are introduced below.

\begin{definition}
	\label{definition 5}
	\textbf{(Difference set).} For $t_1, t_2 \in r$, their difference set is $D(t_1,\, t_2) = \left \{ A \in R \mid t_1[A] \neq t_2[A] \right \} $, representing the set of attributes on which two tuples have different values.
\end{definition}

The difference sets of $r$ is denoted as $D_r = \{D(t_1,\, t_2) \mid t_1,\, t_2 \in r,\, D(t_1,\, t_2) \neq \emptyset\}$. For attribute $A \in R$, the difference sets of $r$ modulo $A$ is denoted as $D_{r}^{A} = \{D\setminus\{A\} \mid D \in D_r \wedge A \in D\}$. A difference set $D \in D_r$ is minimal if there does not exist $D' \in D_r$ and $D' \subset D$. $\underline{D_r} = \left\{ D \in D_r \mid D' \in D_r \land D' \subseteq D \Rightarrow D' = D \right\}$, where $\underline{D_r}$ consists of all minimal difference sets of $D_r$. The \textbf{hypergraph} of $r$ is denoted as $\mathcal{H}_r = (R, D_r)$, where $R$ is the vertex set and $D_r$ is the hyperedge set. The \textbf{sub-hypergraph} of attribute $A$ is denoted by $\mathcal{H}_A = (R, D_r^A)$. The minimal hypergraph $\underline{\mathcal{H}_r} = (R, \underline{D_r})$ contains only the minimal hyperedges of $\mathcal{H}_r$.

\begin{definition}
	\label{definition 7}
	\textbf{(Hitting set).} Given a sub-hypergraph $\mathcal{H}_A = (R, D_r^A)$, $X$ is a hitting set of $\mathcal{H}_A$ iff it intersects with every hyperedge in $D_r^A$. If no subset of $X$ is a hitting set, $X$ is a minimal hitting set. $HS(\cdot)$ denotes the set of minimal hitting sets of a hypergraph.
\end{definition}

\begin{property}
	\label{property1} If an FD $X \rightarrow A$ ($A \notin X$) holds on $r$, $X$ must be a hitting set of $\mathcal{H}_A$. Moreover, $X \rightarrow A$ is a minimal and non-trivial FD iff $X$ is a minimal hitting set of $\mathcal{H}_A$. 
\end{property}

\begin{property}
	\label{property2}
	Any hitting set of $\underline{\mathcal{H}_A}$ is also a hitting set of $\mathcal{H}_A$. The complete minimal hitting sets of $\mathcal{H}_A$ are equal to the complete minimal hitting sets of $\underline{\mathcal{H}_A}$.
\end{property}
It suffices to consider $\underline{\mathcal{H}_r}$ when enumerating hitting sets. Actually, $\underline{\mathcal{H}_r}$ is usually much smaller than $\mathcal{H}_r$. In the following sections, both hypergraphs and the sub-hypergraphs refer to their minimal forms.

\begin{problem}
	Given a relation instance $r$ of schema $R$, the incremental FD discovery aims to find the complete set $\mathcal{F}$ of minimal and non-trivial FDs that hold on the dataset after each data update, without re-executing the algorithm.
\end{problem}

\section{EAIFD Algorithm}\label{sec:algorithm}

\begin{figure*}[!htbp]
\small
    \centering    
    \includegraphics[scale = 0.4]{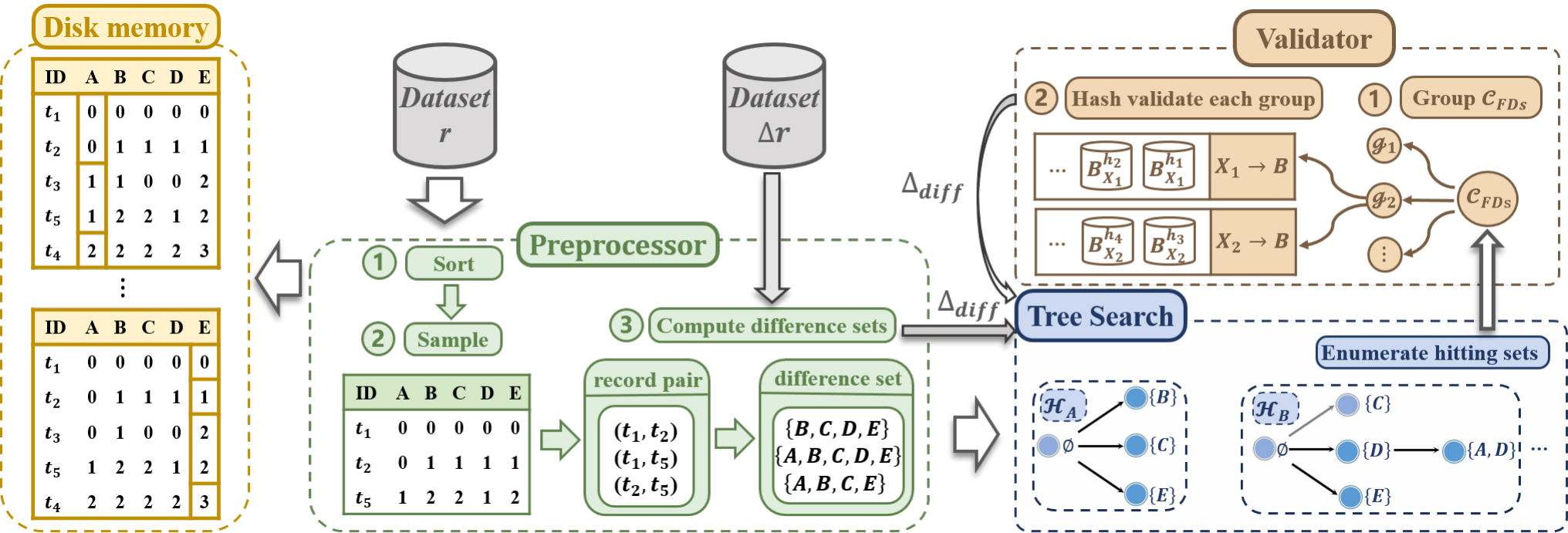}
    \caption{\small Overview of EAIFD and its components.}
    \label{fig:EAIFD}
\end{figure*}

In this section, we introduce EAIFD, a novel FD discovery algorithm designed for incremental databases. EAIFD can efficiently discover the complete set $\mathcal{F}$ of minimal and non-trivial FDs on a dataset and rapidly update $\mathcal{F}$ after each data increment. Compared with existing incremental methods, EAIFD enables fast initialization, scales efficiently on high-dimensional datasets, and reduces memory consumption. The overall framework of EAIFD is illustrated in Fig.~\ref{fig:EAIFD}. 

The core idea of EAIFD is to generate candidate FDs through hitting set enumeration on hypergraphs constructed from difference sets and to validate candidate FDs. The EAIFD is composed of two core components: a \textit{one-time discovery process} for the initial dataset, and an efficient \textit{incremental update mechanism} for all subsequent data increments. The \textit{one-time discovery process} leverages sampling for rapid initialization and builds the crucial multi-attribute hash table ($MHT$) to accelerate future updates. The \textit{incremental update mechanism} intelligently utilizes $MHT$ for rapid validation pruning and employs the IGHV on data blocks for validating remaining candidates efficiently. The details of two components are described in Sections~\ref{subsec:phase1} and~\ref{subsec:phase2}, respectively.

\subsection{One-time discovery process on initial dataset $r$}
\label{subsec:phase1}

\textit{Preprocessing}. The input dataset $r$ is first sorted in ascending order by each attribute and stored on disk, with the index range of each distinct value recorded for rapid loading data blocks in following steps. Then EAIFD applies uniform random sampling without replacement to \textit{r} at a sampling ratio $\varepsilon = 0.3$, selecting $(\binom{n}{2})^{\varepsilon}$ record pairs from total pairs in a dataset with $n$ records. This approach balances data coverage with efficiency and has proved high effectiveness in FD discovery~\cite{DBLP:journals/pacmmod/BleifussPBSN24,DBLP:journals/pvldb/BirnickBFNPS20}. We denote the sampled data as $s$. 

\textit{One-time discovery process}, as listed in Algorithm~\ref{alg:phase1}, generates the complete set $\mathcal{F}$ of minimal non-trivial FDs holding on $r$ and constructs $MHT$ through two core steps: candidate FD generation (Section~\ref{subsubsec:genFDs1}) and validation (Section~\ref{subsubsec:validateFDs1}).

\subsubsection{Generating candidate FDs with sampled data $s$}\label{subsubsec:genFDs1}

EAIFD generates candidate FDs by processing each attribute $A \in R$ as RHS (Lines~\ref{alg:phase1:line2}-\ref{alg:phase1:line27}). For each $A$, it constructs a partial sub-hypergraph $\mathcal{H}_A = (R, D_s^A)$ from sampled data $s$ (Line~\ref{alg:phase1:line2}), where hyperedges in $D_s^A = {D(t_i, t_j) \setminus {A} \mid t_i, t_j \in s \land A \in D(t_i, t_j)}$ represent difference sets of tuple pairs that differ on $A$, with $A$ itself removed. $D_s^A$ captures differences involving $A$ and defines the search space for candidate LHSs with $A$ as the RHS. EAIFD then applies MMCS algorithm for each $\mathcal{H}_A$ to enumerate minimal LHS candidates for $A$ as RHS (Line~\ref{alg:phase1:line6}), forming candidate FDs set $\mathcal{C}_{FDs}$. Here, MMCS~\cite{DBLP:journals/dam/MurakamiU14} is a hyperedge-based branching algorithm for enumerating minimal hitting sets from a hypergraph, which explores the search space efficiently by constructing a search tree and integrating minimality checking to avoid unnecessary branching. Since these candidate FDs are derived from partial sub-hypergraphs based on the sampled dataset $s$, they require validation to ensure validity on $r$.

\begin{algorithm}[!t]
	\caption{EAIFD initial FD discovery}  
	\small
	\label{alg:phase1}
	\DontPrintSemicolon
	\KwIn{Sample data $s$ and frequency threshold $\theta$}
	\KwOut{The complete set of valid FDs $\mathcal{F}$, multi-attribute hash table $MHT$, and sub-hypergraphs $\mathcal{H}$}
	
	\linelabel{alg:phase1:line1} $\mathcal{F}$, $\Delta_{diff}$, $\mathcal{B}$ initialize to empty sets\;
	\linelabel{alg:phase1:line2} Build $\mathcal{H}$ containing sub-hypergraphs for each attribute\;
	
	\linelabel{alg:phase1:line3} \For{$A \in R$}{
		\linelabel{alg:phase1:line4} \Repeat{$\Delta_{diff}$ is empty}{
			\linelabel{alg:phase1:line5} Update $\mathcal{H}$ by $\Delta_{diff}$ and clear $\Delta_{diff}$\;
			\linelabel{alg:phase1:line6} $\mathcal{C}_{FDs} \gets \text{MMCS}(\mathcal{H}_A)$\;
			\linelabel{alg:phase1:line7} $\mathcal{G} \gets \text{Group}(\mathcal{C}_{FDs})$\;
			
			\linelabel{alg:phase1:line8} \ForEach{$g \in \mathcal{G}$}{
				\linelabel{alg:phase1:line9} Select a common attribute $B$ of all LHSs of $g$\;
				
				\linelabel{alg:phase1:line10} \ForEach{block $r_i$ in $r$ sorted by $B$}{
					\linelabel{alg:phase1:line11} \ForEach{tuple $t \in r_i$}{
						\linelabel{alg:phase1:line12} \ForEach{$X_i \to A \in g$}{
							\linelabel{alg:phase1:line13} $h \gets f(\{t[X_i]\})$\;
							\linelabel{alg:phase1:line14} $\mathcal{B}_{X_i}^h \gets \mathcal{B}_{X_i}^h \cup \{t\}$\;
						}
					}
					
					\linelabel{alg:phase1:line15} \ForEach{$\mathcal{B}_{X_i} \in \mathcal{B}$}{
						\linelabel{alg:phase1:line16} \ForEach{$\mathcal{B}_{X_i}^h \in \mathcal{B}_{X_i}$}{
							\linelabel{alg:phase1:line17} \If{$\exists t_1, t_2 \in \mathcal{B}_{X_i}^h \text{ with } t_1[A] \neq t_2[A]$}{
								\linelabel{alg:phase1:line18} Compute difference sets of $(t_1,t_2)$ and add to $\Delta_{diff}$\;
								\linelabel{alg:phase1:line19} $g \gets g \setminus \{X_i \to A\}$\;
								\linelabel{alg:phase1:line20} \textbf{break}\;
							}
						}
					}
					
					\linelabel{alg:phase1:line21} \If{$X_i \to A$ is valid}{
						\linelabel{alg:phase1:line22} \ForEach{$\mathcal{B}_{X_i}^h \in \mathcal{B}_{X_i}$}{
							\linelabel{alg:phase1:line23} \If{the count of tuples in $\mathcal{B}_{X_i}^h \geq \theta \times |r|$}{
								\linelabel{alg:phase1:line24} Add $(h, t[A])$ to $MHT(X \to A)$\;
							}
						}
					}
					
					\linelabel{alg:phase1:line25} Clear $\mathcal{B}_{X_i}$\;
				}
				\linelabel{alg:phase1:line26} $\mathcal{F} \gets \mathcal{F} \cup g$\;
			}
		}
	}
	\linelabel{alg:phase1:line27} \Return{$\mathcal{F},\ MHT,\ \mathcal{H}$}\;
	
\end{algorithm}
\subsubsection{Validating candidate FDs with IGHV}\label{subsubsec:validateFDs1}

\begin{figure*}[!htbp] 
	\small
	\centering
	\includegraphics[scale = 0.45]{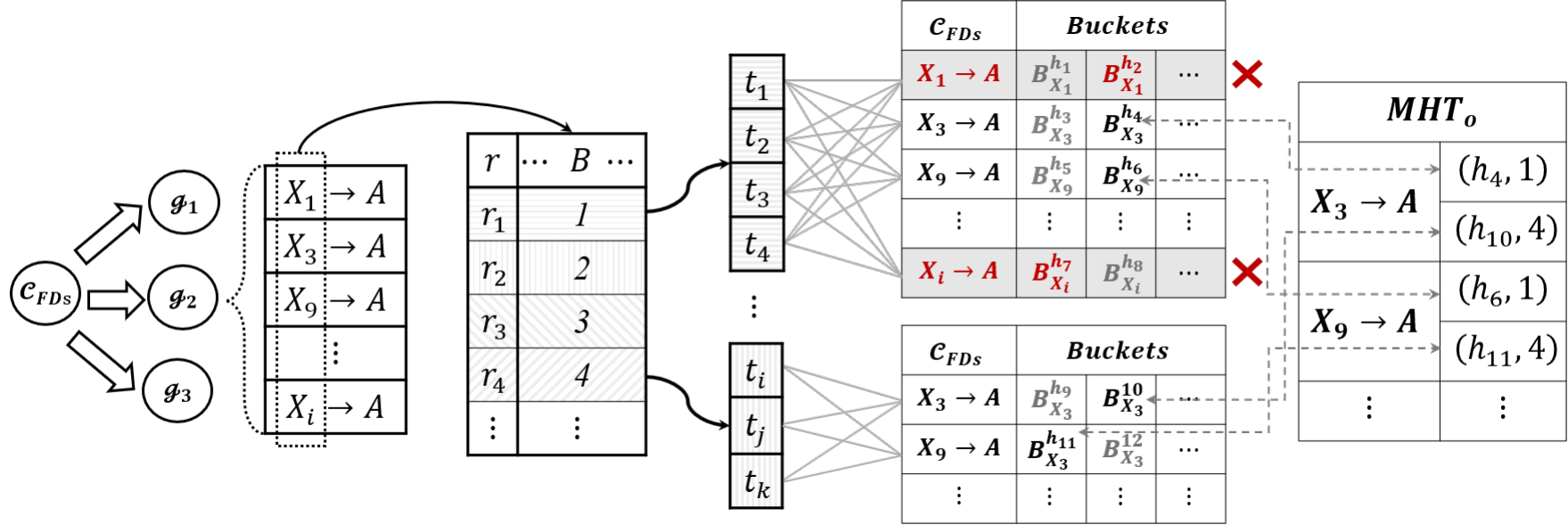}
	\caption{\small The process of IGHV. After grouping, select some $\mathcal{C}_{FDs}$ in $g_2$ and show its core validation steps on data blocks $r_1$ and $r_4$}
	\label{fig:IGHV}
\end{figure*}

We propose an efficient validation method with a low memory usage, called \underline{i}terative \underline{g}rouping \underline{h}ash \underline{v}alidation (IGHV), to check the validity of candidate FDs derived from partial sub-hypergraphs on $r$. Fig.~\ref{fig:IGHV} illustrates its main process. IGHV applies to both static data and incremental updates with few differences. 

To understand the novelty and advantages of IGHV, it is helpful to first review the principles and limitations of a standard hash-based validation approach. Let $\mathcal{B}$ denote the complete set of hash buckets, and $\mathcal{B}_X$ represent the hash bucket set for candidate FD $X \rightarrow A$. $\forall t \in r$, it applies a hash function $f$ to $t[X]$ to compute the hash value $h$, and assigns $t$ to the bucket $\mathcal{B}_X^h$. Any two tuples $t_i, t_j$ in the same bucket $\mathcal{B}_X^h$ satisfy $t_i[X] = t_j[X]$. If $t_i[A] \neq t_j[A]$, a violated tuple pair $(t_i, t_j)$ is detected, and $X \rightarrow A$ is declared invalid. To validate $X \rightarrow A$, this method traverses the entire dataset to build $\mathcal{B}_X$. It works well on small datasets. However, as the number of candidate FDs increases or the dataset scale expands, the memory consumption of hash buckets becomes excessive and may lead to memory overflow.

IGHV significantly improves validation efficiency while effectively controlling memory usage. Algorithm~\ref{alg:phase1} describes the IGHV process for validating candidate FDs $\mathcal{C}_{FDs}$ obtained by the MMCS tree search (Lines~\ref{alg:phase1:line7}-\ref{alg:phase1:line26}). IGHV first groups $\mathcal{C}_{FDs}$ such that the FDs in each group share at least one common LHS attribute (Line~\ref{alg:phase1:line7}). If two candidate FDs $X' \rightarrow A$ and $X'' \rightarrow A$ satisfy $X' \cap X'' \neq \varnothing$, they can be placed in the same group $g$. For each group $g = \{X_1 \rightarrow A, X_2 \rightarrow A, \dots, X_k \rightarrow A\}$, where $X_1 \cap X_2 \cap \dots \cap X_k \neq \varnothing$, a common attribute with uniform value distribution $B \in X_1 \cap X_2 \cap \dots \cap X_k$ is selected as a common sort attribute (Line~\ref{alg:phase1:line9}). This ensures balanced data block sizes, thereby avoiding excessive memory consumption from overly large blocks and frequent loading overhead from too many small blocks. During preprocessing, data is sorted by $B$ and can be divided into blocks $r = r_1 \cup r_2 \cup \dots \cup r_m$, where all tuples in a block $r_i$ share the same value for $B$. During validation, the dataset is loaded in the order sorted by $B$. Each block $r_i$ is processed as follows (Lines~\ref{alg:phase1:line10}-\ref{alg:phase1:line25}): (\romannumeral1) For each tuple $t \in r_i$, IGHV computes the hash values $f(t[X_1]), f(t[X_2]), \dots, f(t[X_k])$ of the LHSs of all candidate FDs in the current group $g$, and inserts them into the corresponding hash bucket sets $\mathcal{B}_{X_1}, \mathcal{B}_{X_2}, \dots, \mathcal{B}_{X_k}$ (Lines~\ref{alg:phase1:line11}-\ref{alg:phase1:line14}). (\romannumeral2) For each candidate FD $X \rightarrow A \in g$, it checks each bucket in the hash bucket set $\mathcal{B}_{X}$ to validate if all tuples in the bucket have the same value on attribute $A$ (Lines~\ref{alg:phase1:line15}-\ref{alg:phase1:line20}). If a hash bucket contains tuples with different values on $A$, a violated tuple pair $(t_p, t_q)$ is identified, where $t_p[X] = t_q[X]$ but $t_p[A] \neq t_q[A]$, indicating that $X \to A$ does not hold. The difference set of these violated tuple pairs is added to the new hyperedge set $\Delta_{diff}$ for subsequent hypergraph updates. To avoid redundant validation, the invalid candidate FD is then removed (Lines~\ref{alg:phase1:line17}-\ref{alg:phase1:line19}). 

After validating data block $r_i$, IGHV clears the corresponding hash buckets (Line~\ref{alg:phase1:line25}) and proceeds to validate the next block. Once all data blocks have been processed, the validation for this group ends, the valid FDs from the group are added to $\mathcal{F}$ (Line~\ref{alg:phase1:line26}). Before validating the next group, the newly collected difference set $\Delta_{diff}$ is used to update the partial sub-hypergraphs in $\mathcal{H}$ corresponding to the difference sets (Line~\ref{alg:phase1:line5}). MMCS then continues to generate new candidate FDs, which undergo the same validation process until no new difference sets are produced. Finally, the same procedure is applied to the remaining attribute sub-hypergraphs.

\subsubsection{Constructing multi-attribute hash table}\label{buildMHT}

During \textit{one-time discovery process}, we introduce multi-attribute hash table ($MHT$), which keeps LHS to RHS value mappings of valid FDs, and integrates with IGHV to prune the search space and accelerate candidate FD validation in incremental scenarios.

The $MHT$ is designed for FD validation and conflict detection, based on the principle for a valid FD $X \rightarrow A$, all tuples agreeing on $X$ must also agree on $A$. For each tuple $t$ in $r$, a mapping item $(f(t[X]), t[A])$ is formed by applying a hash function $f$ to the LHS projection $t[X]$, generating a hash key which is then paired with the RHS value $t[A]$. This approach reduces memory usage and speeds up lookups. To ensure correctness, the hash function $f$ ensures that the hash key uniquely identifies an equivalence class of tuples with respect to $X$, for any tuples $t_i, t_j \in r$ if $t_i[X] = t_j[X]$, then $f(t_i[X]) = f(t_j[X])$; and if $t_i[X] \neq t_j[X]$, then $f(t_i[X]) \neq f(t_j[X])$. The complete $MHT$ is built by computing mapping items for all tuples, clearly reflecting the value mapping from LHS to RHS of valid FDs. If $X \rightarrow A$ holds, each hash key $f(t_i[X])$ must correspond to a unique value $t[A]$. $MHT$ allows fast conflict detection, if a hash key maps to distinct values, indicating a violated tuple pair $t_i, t_j$ such that $t_i[X] = t_j[X]$ but $t_i[A] \neq t_j[A]$. In validation phase, the $MHT$ supports $O(k)$ time complexity, making it particularly efficient for FD validation. 

For large datasets with many candidate FDs, the number of mapping items in $MHT$ may grow rapidly, leading to high memory usage. For this reason, EAIFD employs a \textit{high-frequency mapping items preservation strategy}, caching only the mapping items whose frequency in the dataset exceeds a frequency threshold $\theta$. For any valid FD $X \rightarrow A$, IGHV counts the mapping frequency between $X$ and $A$ (Lines~\ref{alg:phase1:line21}-\ref{alg:phase1:line24}). A hash mapping item $(h, a)$ is cached in $MHT(X \rightarrow A)$ if it satisfies the frequency condition: $\frac{|{t \in r_j \mid r_j \in r, f(t[X]) = h, t[A] = a}|}{|r|} \geq \theta$. The frequent mappings represent commonly observed value associations between LHS and RHS. 

The choice of the threshold $\theta$ must balance a high threshold, which creates a sparse $MHT$ with a weak pruning effectiveness, against a low threshold, which results in an oversized $MHT$ and excessive memory use. In this paper, we set $\theta = 80\%$ based on the three considerations. First, adopting a high-frequency threshold to retain only the most significant elements is well-established practice~\cite{DBLP:conf/sigmod/AgrawalIS93}. Second, value mappings with occurrence frequencies above $80\%$ can effectively capture the most representative attribute-value combinations in real-life datasets. Third, the effectiveness of this choice is further confirmed by the experimental results presented later. This strategy allows for efficient pruning of the validation space using high-frequency mapping items while avoiding $MHT$ becoming excessively large, thereby achieving a balance between performance efficiency and memory usage. 

Once all sub-hypergraphs are processed, \textit{one-time discovery process} ends. It derives $\mathcal{F}$ holding on $r$, multi-attribute hash table $MHT$ for each valid FD $X \rightarrow A$ which contains only high frequency value mapping items, and the hypergraph set $\mathcal{H}$ containing sub-hypergraphs for each attribute.

\subsection{Incremental update mechanism on $r \cup \Delta r$}
\label{subsec:phase2}

When the initial dataset $r$ is updated with incremental data $\Delta r$, \textit{incremental update mechanism} is triggered to efficiently update $\mathcal{F}$ and $MHT$ over dataset $r \cup \Delta r$. Similar to \textit{one-time discovery process}, this mechanism also uses an iterative process where candidate FDs are generated by MMCS and validated by a two-step validation strategy. The mechanism iteratively updates hypergraphs until no new candidates are produced, completing the update of $\mathcal{F}$. The complete procedure of \textit{incremental update mechanism} is outlined in Algorithm~\ref{alg:phase2}. 

\subsubsection{Generating candidate FDs with incremental data $\Delta r$}
\label{subsubsec:genFDs2}

After the incremental update, tuples in the incremental data $\Delta r$ are compared pairwise to generate a new difference set $\Delta_{diff}$ (Line~\ref{alg:phase2:line1}). Since the scale of $|\Delta r|$ is typically small, the time cost of this operation remains acceptable. Based on $\Delta_{diff}$, the corresponding sub-hypergraphs are updated (Line~\ref{alg:phase2:line4}), and MMCS tree search is performed on the updated hypergraphs to generate candidate FDs, which is also denoted as $\mathcal{C}_{FDs}$ (Line~\ref{alg:phase2:line5}). MMCS searches from the tree nodes corresponding to the minimal hitting sets of the hypergraphs constructed from the initial dataset $r$ obtained in the \textit{one-time discovery process} to avoid a full restart. MMCS first updates the hypergraphs by $\Delta_{diff}$ as new hyperedges, and identifies the affected initial minimal hitting sets that fail to cover these new hyperedges. Subsequently, for each affected initial minimal hitting set, MMCS selects candidate attributes from uncovered attributes and extends it to generate new candidate FDs.

However, the candidate FDs $\mathcal{C}_{FDs}$ obtained through the MMCS are only guaranteed to hold locally on $r$ and $\Delta r$ but not on $r \cup \Delta r$. This is because the hypergraphs are constructed solely from the difference sets generated from $r$ during the \textit{one-time discovery process} and from pairwise comparisons within incremental data $\Delta r$, and exclude difference sets derived from tuple pairs between $r$ and $\Delta r$. A candidate FD $X \rightarrow A$ from MMCS, valid on $r$ and $\Delta r$ separately, but may be invalid on $r \cup \Delta r$ if tuples $t_i \in r$ and $t_j \in \Delta r$ satisfy $t_i[X] = t_j[X]$ but $t_i[A] \neq t_j[A]$. Therefore, each candidate FD in $\mathcal{C}_{FDs}$ must undergo further validation to confirm its validity on $r \cup \Delta r$.

The candidate FD $\mathcal{C}_{FDs}$ must hold on $r$ due to two reasons. First, the \textit{one-time discovery process} iteratively expands partial hypergraphs until their minimal hitting sets coincide with those of the global hypergraph constructed from all minimal difference sets of $r$, ensuring that $\mathcal{F}$ is complete for all minimal non-trivial FDs on $r$. It is demonstrated that, during this iterative expansion, each minimal hitting set of the partial hypergraph is either a minimal hitting set of the global hypergraph, or it can be extended into one~\cite{DBLP:conf/edbt/SchirmerP0NHMN19}. Once the expansion yields no new hyperedges, the sets of minimal hitting sets in the partial and global hypergraphs are same, indicating that all minimal non-trivial FDs on $r$ have been captured in $\mathcal{F}$. Second, during incremental updates, the hypergraphs are dynamically updated with difference sets $\Delta_{diff}$ from $\Delta r$. Unless $\Delta_{diff}$ introduces new minimal hyperedges, the original minimal hitting sets remain unchanged. It is proven~\cite{DBLP:conf/icde/XiaoYTMW22} that each new minimal hitting set in the updated hypergraph must contain a previous one, implying it is a specialization of the previous FD. This ensures that candidate FDs generated from updated hypergraphs only exclude those in $\mathcal{F}$ invalidated by $\Delta r$, while all remaining and new candidates still hold on $r$.

To improve validation efficiency, \textit{incremental update mechanism} divides $\mathcal{C}_{FDs}$ into two categories: (\romannumeral1) minimal FDs that hold on $r$, denoted as $\mathcal{C}_1 = \mathcal{C}_{FDs} \cap \mathcal{F}$; and (\romannumeral2) non-minimal FDs that also hold on $r$, denoted as $\mathcal{C}_2 = \mathcal{C}_{FDs} \setminus \mathcal{C}_1$, which can be regarded as specializations of $\mathcal{C}_1$. Two categories are validated separately based on the availability of $MHT$ (Lines~\ref{alg:phase2:line8}-\ref{alg:phase2:line9}). The distinct validation mechanisms for $\mathcal{C}_1$ and $\mathcal{C}_2$ are detailed in Section~\ref{subsubsec:validateType1} and Section~\ref{subsubsec:validateType2}, outlined in Algorithm~\ref{alg:VCF}. 

\begin{figure}[!htbp]               
	\centering
	\includegraphics[scale=0.45]{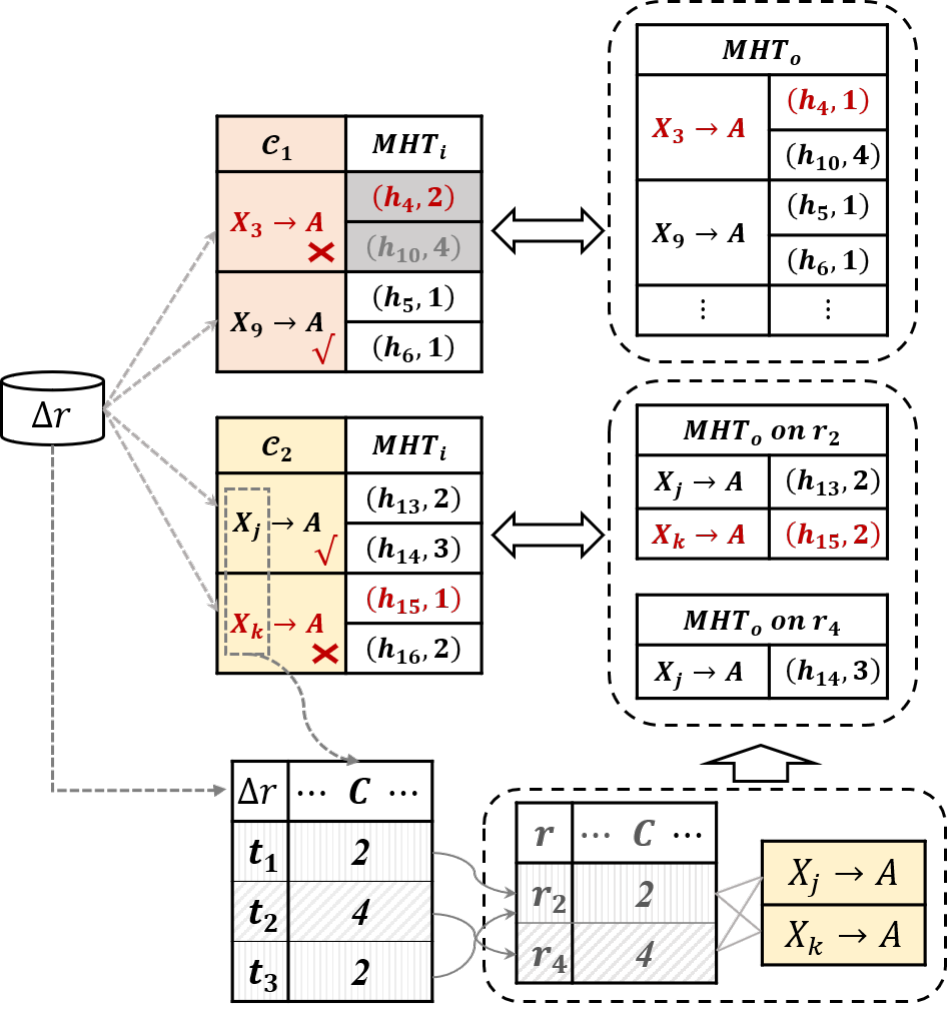}
	\caption{\small The process of validating $\mathcal{C}_{FDs}$ in two types: $\mathcal{C}_1$ and $\mathcal{C}_2$. For clarity, $\Delta r$ contains only 3 tuples, and two candidate FDs are validated in each type.}
	\label{fig:IGHV2}
\end{figure}

\begin{algorithm}[!t]
	\caption{EAIFD incremental FD discovery}
	\small
	\label{alg:phase2}
	\DontPrintSemicolon
	\SetAlgoNlRelativeSize{-1} % 调整行号大小
	\SetNlSty{}{}{:} % 行号样式
	\SetAlgoNlRelativeSize{0}
	
	\KwIn{$\Delta r$, $\mathcal{F}$, $MHT$, $\mathcal{H}$}
	\KwOut{$\mathcal{F}$, $MHT$}
	
	\linelabel{alg:phase2:line1} Compute difference sets of $\Delta r$ and add to $\Delta_{diff}$\;
	
	\linelabel{alg:phase2:line2} \For{$A \in R$}{
		\linelabel{alg:phase2:line3} \While{$\Delta_{diff} \neq \emptyset$}{
			\linelabel{alg:phase2:line4} Update $\mathcal{H}$ by $\Delta_{diff}$ and then clear $\Delta_{diff}$\;
			\linelabel{alg:phase2:line5} $\mathcal{C}_{FDs} \gets \text{MMCS}(\mathcal{H}_A)$\;
			\linelabel{alg:phase2:line6} $\mathcal{C}_1 \gets \mathcal{C}_{FDs} \cap \mathcal{F}\;$\;
			\linelabel{alg:phase2:line7} $\mathcal{C}_2 \gets \mathcal{C}_{FDs} \setminus \mathcal{C}_1$\;
			\tcp{validate candidate FDs in $\mathcal{C}_1$}
			\linelabel{alg:phase2:line8} $\text{res1} \gets \text{VCF}(\mathcal{C}_1,\text{``}{\mathcal{C}_{1}}\_MODE\text{''})$\; 
			\tcp{validate candidate FDs in $\mathcal{C}_2$}
			\linelabel{alg:phase2:line9} $\text{res2} \gets \text{VCF}(\mathcal{C}_2,\text{``}{\mathcal{C}_{2}}\_MODE\text{''})$\;
			\linelabel{alg:phase2:line10} Update $\mathcal{F}$, $\Delta_{diff}$, $MHT$ by res1 and res2\;
		}
	}
	\linelabel{alg:phase2:line13} \Return{$\mathcal{F}$, $MHT$}\;
\end{algorithm}

\subsubsection{Validating candidate FDs in $\mathcal{C}_1$ with IGHV and $MHT$}\label{subsubsec:validateType1}

Since $\Delta r$ is typically much smaller than $r$, most updates only affect the validity of a small portion of FDs. Most tuples still support the original FDs, and most FDs still remain valid after updates. As a result, $\mathcal{C}_1$ accounts for a high proportion of the total candidate FDs. Therefore, improving the validation efficiency of $\mathcal{C}_1$ has become key to optimizing the overall efficiency.

To efficiently validate the candidates in $\mathcal{C}_1$ over $r \cup \Delta r$, EAIFD adopts a two-step validation strategy. The first step,  \textit{$MHT$-based validation}, compares the local hash table built on $\Delta r$ (denoted as $MHT_{\Delta}$) with the pre-computed $MHT$ on $r$ to rapidly identify conflicts and efficiently prune the validation space. Any candidates that cannot be resolved in \textit{MHT-based validation} step proceed to the second step, \textit{table-scan validation}, which loads the relevant data blocks from $r$ and applies IGHV for validation. Figure~\ref{fig:IGHV2} illustrates how EAIFD processes candidate FDs in $\mathcal{C}_1$.

\textit{Step $1$: $MHT$-based validation.} For a candidate FD $X \rightarrow A$, EAIFD first builds a local multi-attribute mapping hash table on $\Delta r$, denoted as $MHT_{\Delta}(X \rightarrow A)$ (Line~\ref{alg:VCF:line3}). Since $|\Delta r|$ is small, the memory usage for $MHT_{\Delta}(X \rightarrow A)$ is manageable. After constructing $MHT_{\Delta}$, \textit{$MHT$-based validation} compares the mapping items in $MHT_{\Delta}(X \rightarrow A)$ with $MHT(X \rightarrow A)$ built on $r$ (Line~\ref{alg:VCF:line5}). During the comparison, the validation result for a candidate FD $X \rightarrow A$ must be one of the following three cases.

\begin{itemize}
	\item \textit{Valid.} If $MHT_{\Delta}(X \rightarrow A) \subseteq MHT(X \rightarrow A)$, all mappings in $MHT_{\Delta}$ already exist in $MHT$, meaning $\Delta r$ does not introduce new LHS to RHS value mappings. $X \rightarrow A$ naturally holds on $r \cup \Delta r$ and is updated to $\mathcal{F}$, and is removed from $\mathcal{C}_1$ without further validation.
	
	\item \textit{Invalid.} If an item $(f(t_j[X]), t_j[A])$ in $MHT_{\Delta}(X \rightarrow A)$ and an item $(f(t_i[X]), t_i[A])$ in $MHT(X \rightarrow A)$ satisfy $f(t_i[X]) = f(t_j[X])$ but $t_i[A] \neq t_j[A]$, then tuples $t_j \in \Delta r$ and $t_i \in r$ form a violated tuple pair. This pair shares the same hash key but has different RHS values, thus $X \rightarrow A$ is invalid on $r \cup \Delta r$ and removed from $\mathcal{C}_1$. The comparison of $(t_i, t_j)$ yields a new difference set, which is added to $\Delta_{diff}$ and used to update the relevant sub-hypergraphs in $\mathcal{H}$ before the next iteration.
	
	\item \textit{Uncertain.} During the comparison between $MHT_{\Delta}(X \rightarrow A)$ and $MHT(X \rightarrow A)$, any mapping item $(h,a)$ in $MHT_{\Delta}(X \rightarrow A)$ that also exists in $MHT$ is removed to reduce redundancy. If $MHT_{\Delta}(X \rightarrow A) \neq \varnothing$ after comparison, it indicates that $MHT_{\Delta}(X \rightarrow A)$ contains mapping items that do not appear in $MHT(X \rightarrow A)$. Such items occur for two reasons: (\romannumeral 1) they exist in the initial dataset $r$ but their occurrence frequency does not meet the frequency threshold $\theta$, and thus are not recorded in $MHT$; or (\romannumeral 2) they represent new LHS to RHS value mappings introduced by $\Delta r$. In this case, the validity of $X \rightarrow A$ remains uncertain and triggers the \textit{table-scan validation}.
\end{itemize}

\textit{Step $2$: table-scan validation.} After the \textit{MHT-based validation}, only candidate FDs classified as \textit{Uncertain} require further table-scan validation, which employs IGHV on the selectively loaded data blocks from the initial dataset. 

Specifically, the \textit{table-scan validation} selectively loads corresponding data blocks from $r$ based on LHS attribute values in $\Delta r$. We only need to detect potential conflicts between a tuple from $\Delta r$ and a tuple from $r$ to check candidate FDs. Conflicts may occur when tuples in $r$ contain the same LHS attribute values as tuples in $\Delta r$, generating same hash keys while having different RHS values. Since tuples within blocks of \textit{r} that do not share the sort attribute value with any tuple in $\Delta r$ cannot produce same hash keys with $\Delta r$ tuples, no conflicts can be detected from them. Scanning such blocks would only incur redundant I/O and computational costs. Since $|\Delta r|$ is typically small, the number of blocks to be loaded is also limited, making this strategy highly effective in improving validation efficiency while ensuring correctness.

\textit{Table-scan validation} first builds local $MHT_{\Delta}$ for each candidate FD on $\Delta r$ (Line~\ref{alg:VCF:line3}). Then, the candidate FDs are grouped using the same grouping strategy as employed in the IGHV of the \textit{one-time discovery process} (Line~\ref{alg:VCF:line7}). After grouping, a sort attribute $B$ is selected for each group (Line~\ref{alg:VCF:line9}).When validating the candidate FDs in this group, EAIFD only loads data blocks from $r$ sorted by $B$ and whose values of $B$ also exist in $\Delta r$. Formally, the set of data blocks to be loaded is defined as:
$r' = \{ r_i \mid t_i \in r_i \wedge t_i[B] \in \{ t_j[B] \mid t_j \in \Delta r, f(t_j[X]) \in K_{\Delta} \} \}$
where $K_{\Delta} = \text{KeySet}(MHT_{\Delta}(X \rightarrow A))$ (Line~\ref{alg:VCF:line10}). The method then validates this group candidates on blocks in $r'$ instead of the full $r$, sequentially performing batch validation on each block.

After obtaining relevant data block set $r'$, EAIFD sequentially loads each data block and performs batch validation on the current block. For each candidate FD $X \to A$ in the group, it first builds a local temporary $MHT(X \rightarrow A)$ to record the mappings from LHS hash key value to RHS value based on the current data block tuples; then, it compares the local temporary $MHT(X \rightarrow A)$ and $MHT_{\Delta}(X \rightarrow A)$ (Line~\ref{alg:VCF:line13}). $X \to A$ is valid only if no conflict is found across all blocks in $r'$. Conversely, if there exist $t_j \in \Delta r$ and $t_i \in r$ satisfying $f(t_i[X]) = f(t_j[X])$ but $t_i[A] \neq t_j[A]$, the $X \to A$ is judged invalid, and the difference set of $(t_i, t_j)$ is added to $\Delta_{diff}$ for hypergraph updating in the next iteration. After each block validation, mapping items in the local temporary $MHT$ with frequency exceeding $\theta$ are recorded and updated to the global complete $MHT$, and then local temporary $MHT$ is cleared to reduce memory usage. After processing all blocks in $r'$, the valid FDs and the new difference set are recorded (Line~\ref{alg:VCF:line14}). The procedure repeats for all groups (Lines~\ref{alg:VCF:line8}-\ref{alg:VCF:line14}). After processing all groups, the validation concludes by returning the final ${Valid}_{FDs}$ and $\Delta_{diff}$ (Line~\ref{alg:VCF:line15}).

Thus, through \textit{$MHT$-based comparison} between $MHT_{\Delta}$ and $MHT$, followed by selective \textit{table-scan validation} using IGHV on loaded data blocks, EAIFD efficiently validates $\mathcal{C}_1$ and identifies minimal FDs that remain valid.

\begin{algorithm}[!t]
	\caption{Validate Candidate FDs (VCF)}
	\small
	\label{alg:VCF}
	\DontPrintSemicolon
	\KwIn{Candidate set $\mathcal{C}$ and validation mode $mode$}
	\KwOut{${Valid}_{FDs}$, $\Delta_{diff}$}
	\tcp{When $mode$ equals "${\mathcal{C}_{1}}\_MODE$": $\mathcal{C} = \mathcal{C}_1$}
	\tcp{When $mode$ equals "${\mathcal{C}_{2}}\_MODE$": $\mathcal{C} = \mathcal{C}_2$}
	\linelabel{alg:VCF:line1} ${Valid}_{FDs}$, $\Delta_{diff}$ initialize to empty sets\;
	
	\linelabel{alg:VCF:line2} \For{each $X \to A \in \mathcal{C}$}{
		\linelabel{alg:VCF:line3} Build $MHT_{\Delta}(X \to A)$ on $\Delta r$\;
		\linelabel{alg:VCF:line4} \If{$mode$ equals $\text{``}{\mathcal{C}_{1}}\_MODE\text{''}$}{
			\linelabel{alg:VCF:line5} Check $MHT_{\Delta}(X \to A)$ and $MHT(X \to A)$ and collect result\;
		}
	}
	
	\linelabel{alg:VCF:line6} Update ${Valid}_{FDs}$, $\mathcal{C}$, $\Delta_{diff}$ by result\;
	\linelabel{alg:VCF:line7} $\mathcal{G} \gets \text{Group}(\mathcal{C})$\;
	
	\linelabel{alg:VCF:line8} \For{each $g \in \mathcal{G}$}{
		\linelabel{alg:VCF:line9} Select a common attribute $B$ of all LHSs of $g$\;
		\linelabel{alg:VCF:line10} \For{each block $r_i \in r'$}{
			\linelabel{alg:VCF:line11} \For{each $X \to A \in g$}{
				\linelabel{alg:VCF:line12} Build $MHT(X \to A)$ on $r_i$\;
				\linelabel{alg:VCF:line13} Check $MHT_{\Delta}(X\to A)$ and $MHT(X \to A)$ and collect result\;
			}
		}
		\linelabel{alg:VCF:line14} Update ${Valid}_{FDs}$, $\Delta_{diff}$ by compare result\;
	}
	
	\linelabel{alg:VCF:line15} \Return{${Valid}_{FDs},  \Delta_{diff}$}\;
	
\end{algorithm}

\subsubsection{Validating candidate FDs in $\mathcal{C}_2$ with IGHV}\label{subsubsec:validateType2}

For candidate FDs in $\mathcal{C}_2$, which hold on $r$ but are non-minimal, no corresponding mapping items exist in $MHT$ since it only stores value mappings for minimal and non-trivial FDs. EAIFD validates them using the same procedure as the \textit{table-scan validation} described in Section~\ref{subsubsec:validateType1} for the uncertain candidate FDs in $\mathcal{C}_1$, loading data blocks from $r$ containing attribute values present in $\Delta r$. This approach ensures correctness while avoiding redundant computation. The process is illustrated in Fig.~\ref{fig:IGHV2}.

After validating $\mathcal{C}_1$ and $\mathcal{C}_2$, EAIFD updates FDs that hold on $r \cup \Delta r$ into $\mathcal{F}$ and returns a difference set $\Delta_{diff}$ from the violated tuple pairs. This $\Delta_{diff}$ updates the relevant hypergraphs in $\mathcal{H}$ for the next iteration.

\subsection{Algorithm analysis}\label{subsec:algorithmAnalysis}

To establish the theoretical soundness and efficiency of EAIFD algorithm, this part first provides a formal proof of correctness and then a detailed complexity analysis. The former guarantees the accuracy of the algorithm, while the latter quantifies its performance advantages.

\subsubsection{Correctness Proof}\label{subsubsec:correctProof}

EAIFD involves candidate generation and validation in both \textit{one-time discovery process} and \textit{incremental update mechanism}, with overall correctness relying on two steps. For candidate generation, its correctness is guaranteed in three aspects: (\romannumeral1) completeness, (\romannumeral2) minimality,  and (\romannumeral3) non-triviality.

\textbf{Completeness.} For each RHS attribute $A$, the \textit{one-time discovery process} builds a sub-hypergraph $\mathcal{H}_A$ and iteratively updates it with new hyperedges derived from violated tuple pairs until no new hyperedges can be generated. As proved by Lemma~\ref{lemma:completeness}, Corollaries~\ref{corollary:validity} and~\ref{corollary:completeness}, when no new hyperedges are generated, $\mathcal{H}_A$ has same minimal hitting sets as the global hypergraph constructed from all minimal difference sets in $r$. Hence, $\mathcal{H}_A$ guarantees complete candidate FD generation. MMCS enumerates all minimal hitting sets of $\mathcal{H}_A$, each corresponding to a valid minimal FD $X \rightarrow A$ per Definition~\ref{definition 7}.

Completeness in the \textit{incremental update mechanism} relies on the same principle of iterative hypergraph refinement used in the \textit{one-time discovery process}. This mechanism starts with $\mathcal{H}_A$, upon which completeness in \textit{one-time discovery process} is proven, and updates it iteratively by adding two types of new difference sets: those from pairwise comparisons within $\Delta r$, and those generated by violated tuple pairs between $\Delta r$ and $r$. Once no new hyperedges appear, Lemma~\ref{lemma:completeness} and Corollaries~\ref{corollary:validity}–\ref{corollary:completeness} ensure that MMCS enumerates the same minimal hitting sets as those derived from all minimal difference sets over $r \cup \Delta r$, thereby guaranteeing completeness in candidate FD generation in the \textit{incremental update mechanism}.

\begin{lemma}
	\label{lemma:completeness}
	For any two hypergraphs $\mathcal{H}$ and $\mathcal{H'}$, $\mathcal{H} \prec \mathcal{H'}$ if and only if $HS(\mathcal{H}) \supseteq HS(\mathcal{H'})$.~\cite{DBLP:journals/pvldb/BirnickBFNPS20}
\end{lemma}

\begin{corollary}
	\label{corollary:validity}
	A minimal hitting set $X$ for the partial hypergraph $\mathcal{P}_A$ is a global minimal FD (i.e., $X \in HS(\mathcal{D}_A)$) if and only if the FD $X \to A$ holds on the dataset.~\cite{DBLP:journals/pvldb/BirnickBFNPS20}
\end{corollary}

\begin{corollary}
	\label{corollary:completeness}
	When the algorithm iteration produces no new difference sets (i.e., new hyperedges), the minimal hitting sets $HS(\mathcal{P}_A)$ of the current hypergraph $\mathcal{P}_A$ constitute all minimal FDs, i.e., $HS(\mathcal{P}_A) = HS(\mathcal{D}_A)$.~\cite{DBLP:journals/pvldb/BirnickBFNPS20}
\end{corollary}

\textbf{Minimality.} As stated in Property~\ref{property1}, MMCS guarantees the minimality of candidate FDs in both the \textit{one-time discovery process} and the \textit{incremental update mechanism}, since it enumerates only minimal hitting sets. Each hitting set $X$ on $\mathcal{H}_A$ directly corresponds to a minimal FD $X \rightarrow A$.

\textbf{Non-triviality.} To ensure non-triviality, \textit{one-time discovery process} constructs sub-hypergraph $\mathcal{H}_A$ using difference sets generated from tuple pairs different on $A$, while excluding $A$. \textit{One-time discovery process} and the \textit{incremental update mechanism} update $\mathcal{H}_A$ with new difference sets generated from tuple pairs different on $A$, also excluding $A$.
MMCS then enumerates hitting sets on $\mathcal{H}_A$.
According to Definitions~\ref{definition 5} and~\ref{definition 7}, since $A \notin X$ and $X$ determines $A$, the FD $X \rightarrow A$ is non-trivial.

The second essential step ensuring correctness of EAIFD is candidate validation, whose validation strategies in both phases accurately determine the validity of candidates and identify all violated tuple pairs necessary for hypergraph refinement.

\textit{One-time discovery process} employs IGHV to validate candidate FDs. IGHV first groups the candidates and selects a common sort attribute for each group. It then sequentially loads data blocks sorted by this attribute and performs validation within each block. Under the hash function $f$, tuples with identical LHS values share the same hash value, while those with different LHS values have distinct hashes, ensuring that tuples with equal hashes are located in the same block after sorting. Consequently, validation only needs to examine tuple pairs within the current block for possible violations, which localizes comparisons and greatly reduces cost while preserving correctness. If any block contains violated tuple pairs, the FD is deemed invalid; IGHV generates difference sets to update the hypergraphs. If no conflicts are found across all blocks, the FD is considered valid.

\textit{Incremental update mechanism} ensures validation correctness by a two-step validation strategy. Candidate FDs are divided into $\mathcal{C}_1$ and $\mathcal{C}_2$ and validated through distinct methods. Candidates in $\mathcal{C}_1$ are rapidly validated by the \textit{MHT-based validation} through comparing items in $MHT_{\Delta}$ with the items in $MHT$. Since $MHT$ stores the high-frequency hash LHS value to RHS value mapping items of candidate FDs from $r$ and $MHT_{\Delta}$ stores all hash LHS value to RHS value mapping items from $\Delta r$, the comparison between them can accurately detect violated tuple pairs between $\Delta r$ and $r$. The remaining candidate FDs in $\mathcal{C}_1$ that cannot be validated by \textit{MHT-based validation}, along with $\mathcal{C}_2$, are validated using \textit{table-scan validation}, which performs IGHV through selective loading of data blocks. After grouping the candidate FDs and selecting a common sort attribute, this method does not load all data blocks of $r$. Instead, it selectively loads only those blocks containing tuples that share the same value as the tuples in $\Delta r$ on the common sort attribute. Only tuples within these blocks may share the same LHS hash values as the incremental data, thus requiring validation of their RHS attributes to detect violated tuple pairs. Since tuples in other blocks have different sort values and thus distinct LHS hashes, they are safely skipped. This method guarantees correctness while substantially reducing I/O and computation cost.

Therefore, based on the demonstrated correctness of both candidate generation and validation steps, the EAIFD algorithm is proven to be correct.

\subsubsection{Complexity Analysis}

This part comprehensively analyzes the time complexity of EAIFD, covering both \textit{one-time discovery process} and \textit{incremental update mechanism}, and concludes with the analysis of space complexity. We consider a dataset $r$ with $n$ tuples and $m$ attributes, and the incremental data $\Delta r$ with $n_{\Delta}$ tuples.

\textbf{Time complexity.} We begin with the \textit{one-time discovery process}. Its preprocessing involves sorting $r$ by each attribute for subsequent validation, taking $O(m \times n \log n)$ time. Then, tuples in sample data $s$ are compared pairwise to compute the difference sets with complexity $O(|s|^2)$. This avoids the $O(m \times n^2)$ complexity of DHSFD requiring all pairwise tuple comparisons and enables rapid initialization of EAIFD.

Following preprocessing, \textit{one-time discovery process} iteratively performs candidate generation and validation for each RHS attribute $A$. Within each iteration, candidate generation relies on MMCS algorithm. As proved in~\cite{DBLP:journals/dam/MurakamiU14}, the time complexity for a single search step of MMCS on sub-hypergraph $\mathcal{H}_A = (R, D_r^A)$ is $O(|\mathcal{H}_A|)$, where $| \mathcal{H}_A | = \sum_{D \in  D_r^A} |D|$ denotes the total size of hyperedges. Discovering one minimal hitting set typically requires multiple MMCS search steps, with most search steps used only for state updates or pruning. Generating $k$ minimal hitting sets in an iteration takes time complexity of $O(k \times |\mathcal{H}_A| \times \alpha)$, where $\alpha$ is the average number of search steps to discover one minimal hitting set. 

In the same iteration, candidate validation employs IGHV to validate the $k$ candidate FDs generated by MMCS. In the worst case, each candidate FD needs to be validated across all data blocks with complexity $O(k \times n)$, while candidates are pruned immediately upon conflict. Let $b$ be the average number of blocks validated per candidate FD and $|V|$ the distinct values of the common sort attribute in $r$ for the group. With approximate block size $\frac{n}{|V|}$, the validation cost per candidate FD is $O(b \times \frac{n}{|V|})$, yielding average complexity $O(k \times b \times \frac{n}{|V|})$ for $k$ candidates. 

Combining the candidate generation and validation, the time complexity for one iteration is $O(k \times (|\mathcal{H}_A| \times \alpha + b \times \frac{n}{|V|}))$. The total cost of \textit{one-time discovery process} can be obtained by multiplying this one iteration cost by the number of iterations, which is dependent on the underlying data.

Next, we analyze the time complexity of \textit{incremental update mechanism}. This mechanism first computes new difference sets from pairwise comparisons of tuples in $\Delta r$, with time complexity $O(n_{\Delta}^2)$. It then iteratively generates candidate FDs using MMCS and validates them. Since the complexity of MMCS is discussed previously, our focus here is on validation. Candidate FDs are divided into $\mathcal{C}_1$ and $\mathcal{C}_2$, and a two-step validation strategy is applied. We then formally characterize the time complexity of both validation steps for $k$ candidates.

For candidate FDs in $\mathcal{C}_1$, \textit{MHT-based validation} constructs a local $MHT_{\Delta}$ for each $X \rightarrow A$ in $\Delta r$ at a cost of $O(k \times n_{\Delta})$. Comparing $MHT_{\Delta}$ with $MHT$ requires traversing $MHT_{\Delta}$ items and querying $MHT$, which takes $O(1)$ per candidate FD due to hash table lookups, yielding $O(k)$ total for $k$ candidates. The overall complexity for validating $k$ candidates in $\mathcal{C}_1$ is $O(k \times (n_{\Delta} + 1))$, which is highly efficient since $n_{\Delta}$ is typically small.

\textit{Table-scan validation} handles uncertain candidates in $\mathcal{C}_1$ and all candidate FDs in $\mathcal{C}_2$. Assuming a total of $k$ candidates, a local $MHT_{\Delta}$ is first built for each FD on $\Delta r$, costing $O(k \times n_{\Delta})$. Candidates are grouped by common LHS attributes, and each group is validated sequentially. Unlike the \textit{one-time discovery process} that loads all blocks, \textit{table-scan validation} selectively loads only blocks containing attribute values in $\Delta r$. Let $V_{\Delta}$ and $V$ denote the distinct values of the common sort attribute in $\Delta r$ and $r$, respectively. The loaded blocks contain roughly $n \times \frac{|V_{\Delta}|}{|V|}$ tuples. Candidates are pruned immediately upon invalidation. Let $b'$ be the average number of blocks validated per candidate. The total cost of per group containing $\frac{k}{|\mathcal{G}|}$ candidates is $O\left( \frac{k}{|\mathcal{G}|} \times b' \times \frac{n}{|V|} \right)$, giving a total complexity of $O\left(k \times \left( n_{\Delta} + b' \times \frac{n}{|V|} \right)\right)$. Compared to $O(k \times b \times \frac{n}{|V|})$ complexity of validating $k$ candidates using IGHV in the \textit{one-time discovery process}, \textit{table-scan validation} is more efficient since $n_{\Delta} \ll n$ and $b' \ll b$.

\textbf{Space complexity.} EAIFD uses $MHT$ as its core auxiliary structure to accelerate candidate FD validation while controlling memory. Without filtering, each FD could store up to $n$ hash entries, yielding a worst-case space complexity of $O(|\mathcal{F}| \times n)$. To reduce memory, EAIFD applies a \textit{high-frequency mapping preservation} with threshold $\theta \in (0,1)$, caching only hash keys $h$ satisfying $\frac{count(h)}{n} \geq \theta$. Let $p$ denote the cached items per FD; since $p \times (\theta \times n) \le n$, we have $p \le \frac{1}{\theta}$, giving an upper bound of $O(\frac{1}{\theta})$ per FD. Consequently, the overall space complexity is $O(\frac{|\mathcal{F}|}{\theta})$, independent of dataset size $n$ and scalable for large datasets, enabling EAIFD to maintain controllable memory usage even for large-scale datasets.

In summary, EAIFD reduces preprocessing to $O(m \times n \log n)$, avoiding the $O(m \times n^2)$ cost of pairwise comparisons and enabling rapid initialization. Its optimized validation achieves near-linear time in both the \textit{one-time discovery process} and \textit{incremental update mechanism}. Along with the \textit{high-frequency mapping items preservation strategy}, space complexity is bounded by $O(\frac{|\mathcal{F}|}{\theta})$, allowing EAIFD to maintain correctness and scale efficiently on incrementally updated datasets.

\section{Performance evaluation}
\label{sec:performanceEvaluation}
This section evaluates the performance of EAIFD on a DELL OptiPlex Tower Plus 7010 desktop PC equipped with an Intel(R) Core(TM) i7-13700 CPU (16 cores, 3.00 GHz), 32 GB of RAM, running 64-bit Windows 11. All experiments are implemented in Java (JDK 8).

\textbf{Datasets.} We evaluate EAIFD on $20$ different real datasets to assess its practical performance. Table~\ref{tab:experimentResult} summarizes their properties, which differ in tuple number, attribute number, and data distribution, enabling a comprehensive performance analysis. All real datasets are publicly available\footnote{Datasets are available at \url{https://hpi.de/naumann/projects/repeatability/data-profiling/fds.html} and \url{https://www.kaggle.com/datasets}.}.

\textbf{NULL Semantics.} In relational databases, NULL represents missing or inapplicable values, which are common in real-world datasets. Different NULL-handling strategies impact the validity of related FDs. We adopt the NULL-equals-NULL semantics, treating all missing values as equal, which is the common default in FD discovery~\cite{DBLP:journals/pvldb/Berti-EquilleHN18,DBLP:conf/icde/WeiL19}.

\textbf{Competitors.} The static FD discovery algorithm FDHITS~\cite{DBLP:journals/pacmmod/BleifussPBSN24} and two incremental FD discovery algorithms, DynFD~\cite{DBLP:conf/edbt/SchirmerP0NHMN19} and DHSFD~\cite{DBLP:conf/icde/XiaoYTMW22}, are selected for comparison\footnote{All codes are obtained from \url{https://hpi.de/naumann/projects/repeatability/data-profiling/fds.html} and \url{https://github.com/RangerShaw/DHSFD}.}. FDHITS outperforms other static algorithms~\cite{DBLP:journals/pacmmod/BleifussPBSN24}. DynFD (an attribute-based incremental algorithm) maintains positive and negative covers to generate and validate candidate FDs via PLIs, while DHSFD (a tuple-based incremental algorithm) formulates FD discovery as a hitting set enumeration problem over hypergraphs of difference sets with a dynamic enumeration strategy. Comparing these three algorithms with EAIFD provides a comprehensive evaluation of its incremental effectiveness.

\textbf{Parameters.} EAIFD performance is influenced by two parameters: the sampling ratio $\varepsilon$ and the mapping frequency threshold $\theta$. Given a dataset with $n$ records, the total number of record pairs is $p = \binom{n}{2}$. EAIFD uniformly samples approximately $p^\varepsilon$ record pairs ($0 < \varepsilon < 1$) without replacement to build initial partial hypergraphs. Prior studies indicate that setting $\varepsilon=0.3$ achieves a good balance between coverage and efficiency~\cite{DBLP:journals/pvldb/BirnickBFNPS20,DBLP:journals/pacmmod/BleifussPBSN24}. The mapping frequency threshold $\theta$ determines how many items are stored in the $MHT$, thus controlling pruning strength and memory use. Experiments show $\theta=80\%$ provides an effective balance. Thus, we adopt $\varepsilon=0.3$ and $\theta=80\%$ as default settings.

\textbf{Other settings.} From each dataset, we randomly select $80\%$ of tuples as $r$ and generate $\Delta r$ from the remaining $20\%$ as in DHSFD~\cite{DBLP:conf/icde/XiaoYTMW22}. All times are in seconds unless specially stated; TL indicates a run exceeding the limit of 5 hours. Results are averaged over three executions, with the machine restarted each time.

\textbf{Experimental Structure.} 
We conduct comprehensive experiments to evaluate the efficiency, scalability, and unique features of EAIFD. Experiment $1$ on real-life datasets demonstrates the performance advantages of EAIFD. Experiment $2$, $3$ and $4$ analyze scalability across tuple number, attribute number, and incremental data size. Experiment $5$ compares EAIFD against the current state-of-the-art static algorithm in incremental updates. Experiments $6$, $7$ and $8$ respectively demonstrate the key advantages of EAIFD: rapid initialization, a memory-light $MHT$ with \textit{high-frequency mapping items preservation strategy}, and the effect of $\theta$ across a wide range.

\textbf{Exp $1$: Overall performance comparison.} Experiment $1$ evaluates EAIFD against current incremental FD discovery algorithms on $20$ real-world datasets, with $\frac{|\Delta r|}{|r|}$ ranging from $10\%$ to $30\%$, $\varepsilon=0.3$, and $\theta=80\%$. As reported in Table~\ref{tab:experimentResult}. EAIFD is consistently faster than DynFD and DHSFD across all insertion proportions.

Against DynFD, EAIFD achieves speedups of up to two orders of magnitude on several datasets (e.g., Plista). The advantages are particularly significant on larger datasets like Flights, Census, Fd-Reduced-30, and CAB, where DynFD consistently exceeds the time limit (TL). This confirms the scalability limitations of DynFD due to its layer-by-layer candidate generation and PLI-based validation. In contrast, EAIFD transforms candidate generation as a hitting set enumeration problem on hypergraphs and employs a two-step validation strategy, achieving a higher scalability.

Although both DHSFD and EAIFD can handle large-scale datasets, EAIFD runs nearly an order of magnitude faster than DHSFD on several datasets (e.g., Nursery). The significant speedup validates the effectiveness of two-step validation strategy in EAIFD. \textit{MHT-based validation} rapidly processes the majority of candidates ($\mathcal{C}_1$) using efficient $O(1)$ hash lookups. Subsequently, \textit{table-scan validation} selectively loads data blocks and validates the remaining FDs in batch. This strategy substantially improves overall efficiency.

\begin{table*}[]
	\renewcommand{\arraystretch}{1.3}
	\centering
	\caption{\small The experimental results on real-life datasets. Runtime in seconds for EAIFD compared with two incremental algorithms DynFD and DHSFD. The fastest runtimes are highlighted in bold. TL indicates that the algorithm exceeded the time limit.}
	\label{tab:experimentResult}
	\resizebox{\textwidth}{!}{
		\renewcommand{\arraystretch}{1.0}
		\setlength{\tabcolsep}{4pt}
		\begin{tabular}{cccc|ccc|ccc|ccc}
			\hline
			\multicolumn{4}{c|}{Dataset Properties} & \multicolumn{3}{c|}{Incremental $\left ( \left | \Delta r \right | =10\%\left |r\right | \right ) $} & \multicolumn{3}{c|}{Incremental $\left ( \left | \Delta r \right | =20\%\left |r\right | \right ) $} & \multicolumn{3}{c}{Incremental $\left ( \left | \Delta r \right | =30\%\left |r\right | \right ) $} \\ \hline
			\multicolumn{1}{c|}{Dataset} & \multicolumn{1}{c}{$\left | r \right | $} & \multicolumn{1}{c}{$\left | R \right | $} & $|\mathcal{F}|$ & DynFD & DHSFD & EAIFD & DynFD & DHSFD & EAIFD & DynFD & DHSFD & EAIFD \\ \hline
			\multicolumn{1}{c|}{Bridges} & 108 & 13 & 142 & 0.2 & 0.0013 & \textbf{0.001} & 0.23 & 0.004 & \textbf{0.003} & 0.28 & 0.008 & \textbf{0.005} \\
			\multicolumn{1}{c|}{Balance} & 625 & 5 & 1 & 0.13 & 0.03 & \textbf{0.006} & 0.15 & 0.06 & \textbf{0.007} & 0.21 & 0.071 & \textbf{0.009} \\
			\multicolumn{1}{c|}{Abalone} & 4177 & 9 & 137 & 0.1 & 0.03 & \textbf{0.008} & 0.21 & 0.07 & \textbf{0.011} & 0.31 & 0.094 & \textbf{0.024} \\
			\multicolumn{1}{c|}{Iris} & 147 & 5 & 4 & 0.166 & 0.03 & \textbf{0.009} & 0.21 & 0.037 & \textbf{0.009} & 0.2 & 0.049 & \textbf{0.012} \\
			\multicolumn{1}{c|}{Nursery} & 13000 & 9 & 1 & 0.8 & 0.17 & \textbf{0.015} & 1.1 & 0.21 & \textbf{0.018} & 1.6 & 0.235 & \textbf{0.027} \\
			\multicolumn{1}{c|}{NCV} & 1000 & 19 & 758 & 0.3 & 0.04 & \textbf{0.02} & 0.5 & 0.09 & \textbf{0.05} & 0.54 & 0.11 & \textbf{0.07} \\
			\multicolumn{1}{c|}{Hepatitis} & 155 & 20 & 8250 & 1.2 & 0.25 & \textbf{0.07} & 2.5 & 0.3 & \textbf{0.09} & 2.8 & 0.384 & \textbf{0.121} \\
			\multicolumn{1}{c|}{Claim} & 20000 & 11 & 12 & 0.42 & 1.8 & \textbf{0.27} & 0.47 & 2.5 & \textbf{0.45} & 0.7 & 3.5 & \textbf{0.507} \\
			\multicolumn{1}{c|}{Letter} & 20000 & 17 & 61 & 14 & 1.5 & \textbf{0.53} & 22.1 & 1.9 & \textbf{0.77} & 28 & 2.6 & \textbf{1.21} \\
			\multicolumn{1}{c|}{Plista} & 1001 & 63 & 173409 & 143.4 & 4.6 & \textbf{0.5} & 147.5 & 7.5 & \textbf{1.3} & 145.1 & 12.8 & \textbf{3.2} \\
			\multicolumn{1}{c|}{Horse} & 300 & 29 & 128726 & 30.1 & 2.1 & \textbf{0.9} & 33 & 4.1 & \textbf{1.8} & 34.5 & 5.5 & \textbf{3.906} \\
			\multicolumn{1}{c|}{Bioentry} & 184292 & 9 & 19 & 69 & 5.6 & \textbf{2.6} & 83 & 14.8 & \textbf{8.4} & 103 & 34.1 & \textbf{17.9} \\
			\multicolumn{1}{c|}{Tax} & 1000000 & 15 & 263 & 346 & 52 & \textbf{29} & 387 & 93 & \textbf{57} & 494 & 176 & \textbf{87} \\
			\multicolumn{1}{c|}{Ditag\_feature} & 3960124 & 13 & 58 & TL & 358 & \textbf{44} & TL & 1439 & \textbf{287} & TL & 5204 & \textbf{743} \\
			\multicolumn{1}{c|}{Fd-Reduced-30} & 250000 & 30 & 89571 & TL & 144 & \textbf{76} & TL & 367 & \textbf{174} & TL & 583 & \textbf{297} \\
			\multicolumn{1}{c|}{Flight} & 250000 & 31 & 117367 & TL & 190 & \textbf{83} & TL & 410 & \textbf{198} & TL & 630 & \textbf{410} \\
			\multicolumn{1}{c|}{Pitches} & 250000 & 40 & 608928 & TL & 310 & \textbf{132} & TL & 600 & \textbf{420} & TL & 920 & \textbf{780} \\
			\multicolumn{1}{c|}{CAB} & 67300 & 54 & 3353531 & TL & 330 & \textbf{197} & TL & 700 & \textbf{580} & TL & 1350 & \textbf{990} \\
			\multicolumn{1}{c|}{Census} & 196000 & 42 & 41861 & TL & 341 & \textbf{206} & TL & 622 & \textbf{461} & TL & 1077 & \textbf{845} \\
			\multicolumn{1}{c|}{Lineitem} & 6000000 & 16 & 3984 & TL & 1308 & \textbf{968} & TL & 3314 & \textbf{2767} & TL & 7123 & \textbf{6594} \\ \hline
		\end{tabular}
	}
\end{table*}

\begin{figure*}[!htb]               
	\centering
	\includegraphics[scale=0.4]{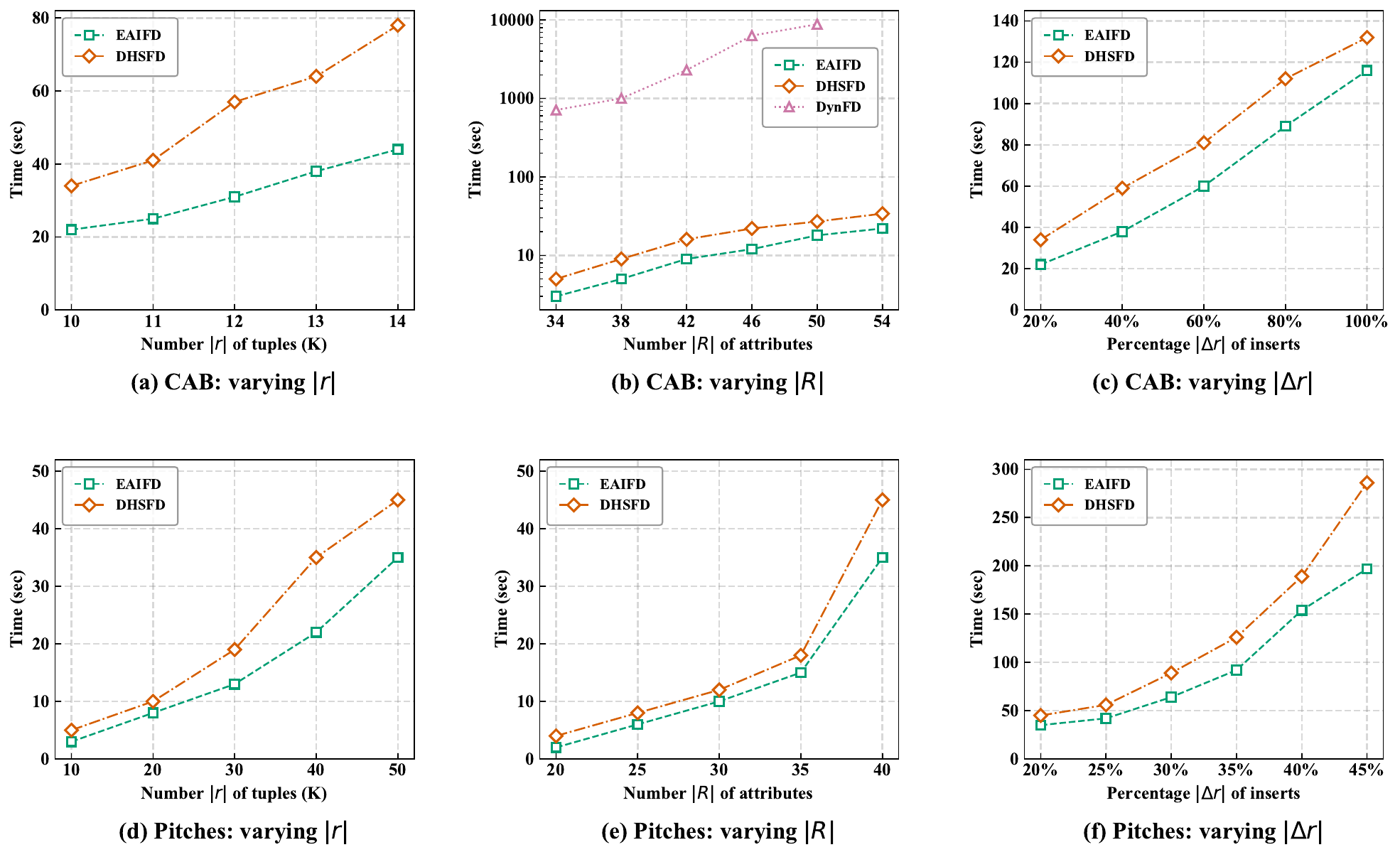}
	\caption{\small Time performance of EAIFD compared with DynFD, and DHSFD on the CAB and Pitches. Subfigures (a, b, c) correspond to CAB with varying tuple number $|r|$, attribute number $|R|$, and incremental data size $|\Delta r|$, respectively, while (d, e, f) show the same for Pitches.}
	\label{fig:exp1}
\end{figure*}

\textbf{Exp $2$: Scalability with tuple number $|r|$.} Experiment~$2$ evaluates the row scalability of EAIFD, DynFD, and DHSFD on the CAB and Pitches datasets by varying $|r|$, as shown in Fig.~\ref{fig:exp1}(a) and Fig.~\ref{fig:exp1}(d). We keep $\frac{|\Delta r|}{|r|}=20\%$ and fix $|R|$ ($54$ for CAB, $40$ for Pitches). Specifically, $|r|$ ranges from $10{,}000$ to $14{,}000$ ($|\Delta r|$ from $2{,}000$ to $2{,}800$) for CAB, and from $10{,}000$ to $50{,}000$ ($|\Delta r|$ from $2{,}000$ to $10{,}000$) for Pitches. Runs exceeding 5 hours are omitted.

DynFD fails to complete within the time limit on both datasets because its PLI-based validation cost grows rapidly with $|r|$, resulting in poor row scalability. In contrast, DHSFD and EAIFD scale more gently with increasing $|r|$, showing stable performance. Notably, EAIFD consistently outperforms DHSFD. The row scalability of DHSFD is limited by the increasing costs associated with updating PLI, computing new difference sets, and refining hypergraphs as $|r|$ increases. By updating the hypergraph through difference sets derived from tuple pairs in $\Delta r$, EAIFD avoids redundant full-pair comparisons involving \textit{r}. Then, two-step validation does not scan the initial dataset but leverages $MHT$ and loads relevant data blocks for validation, ensuring high row scalability and stable performance as $|r|$ increases.

\textbf{Exp $3$: Scalability with attribute number $|R|$.} Experiment~$3$ evaluates the column scalability of EAIFD, DynFD, and DHSFD on the CAB and Pitches datasets by varying $|R|$, with $|r|$ and $|\Delta r|$ fixed ($10{,}000$ and $2{,}000$ for CAB; $50{,}000$ and $10{,}000$ for Pitches). Fig.~\ref{fig:exp1}(b) and Fig.~\ref{fig:exp1}(e) show results when varying $|R|$ from $34$–$54$ for CAB and $20$–$40$ for Pitches, respectively.

DynFD exhibits poor column scalability on both datasets. On CAB, its runtime rises sharply with $|R|$ ($713$s-$8{,}831$s) and fails to finish within the limit at $|R|=54$. On Pitches, DynFD failed to complete within the time limit across all tested attribute numbers ($|R|$ from $20$ to $40$). This severe performance degradation stems from its reliance on PLI-based validation, where increasing attribute numbers lead to a rapid growth in both candidate number and set intersection operations. 

As observed in Fig.~\ref{fig:exp1}(b) and (e), both tuple-based algorithms, DHSFD and EAIFD, indicate significantly better column scalability than column-based DynFD, exhibiting moderate runtime growth as $|R|$ increases. EAIFD consistently outperforms DHSFD, showing lower runtime and growth trend. The performance of DHSFD is still affected by the increasing complexity of updating PLI, difference sets and hypergraphs as more attributes are involved. In contrast, EAIFD achieves a superior column scalability due to two key choices. For one thing, EAIFD prevents excessive candidates growth by modeling FD generation as hitting set enumeration on partial hypergraph and generating new candidates selectively. For another, its two-step validation strategy replaces costly PLI-based set intersections by $MHT$ structure for rapid validation, and then validating the remaining candidate FDs in batches over relevant data blocks, ensuring high column scalability.

\textbf{Exp $4$: Scalability with incremental size $|\Delta r|$.} Experiment $4$ evaluates the incremental scalability of EAIFD, DynFD, and DHSFD on CAB ($|r|=10{,}000$, $|R|=54$, $|\Delta r|=2{,}000$–$10{,}000$, $\frac{|\Delta r|}{r}=20\%-100\%$) and Pitches ($|r|=50{,}000$, $|R|=40$, $|\Delta r|=10{,}000$–$22{,}500$, $\frac{|\Delta r|}{r}=20\%-45\%$) as shown in Fig.~\ref{fig:exp1}(c) and Fig.~\ref{fig:exp1}(f). 

On both datasets, DynFD fails to complete within the time limit, confirming its poor scalability, as its PLI-based validation across $r \cup \Delta r $ becomes prohibitive. The runtimes of DHSFD and EAIFD both grow noticeably as the incremental ratio $\frac{|\Delta r|}{r}$ increases, incurred by higher costs in difference set computation, hypergraph refinement and validation operation. However, EAIFD consistently outperforms DHSFD and exhibits a lower growth rate in runtime. The key advantage for EAIFD stems directly from two aspects. First, EAIFD avoids enumerating all cross-set pairs and updates difference sets within $\Delta r$, while DHSFD computes difference sets between $\Delta r$ and $r$. Second, a two-step validation strategy enables efficient candidate validation. Even when $\frac{|\Delta r|}{r}$ increases, MHT-based validation in EAIFD handles most $\mathcal{C}_1$ candidates efficiently via rapid conflict detection. The subsequent table-scan validation only loads relevant data blocks for the remaining candidates in batch, reducing the unnecessary I/O cost.

\begin{figure}[!htbp]               
	\centering
	\includegraphics[width=0.48\textwidth, keepaspectratio]{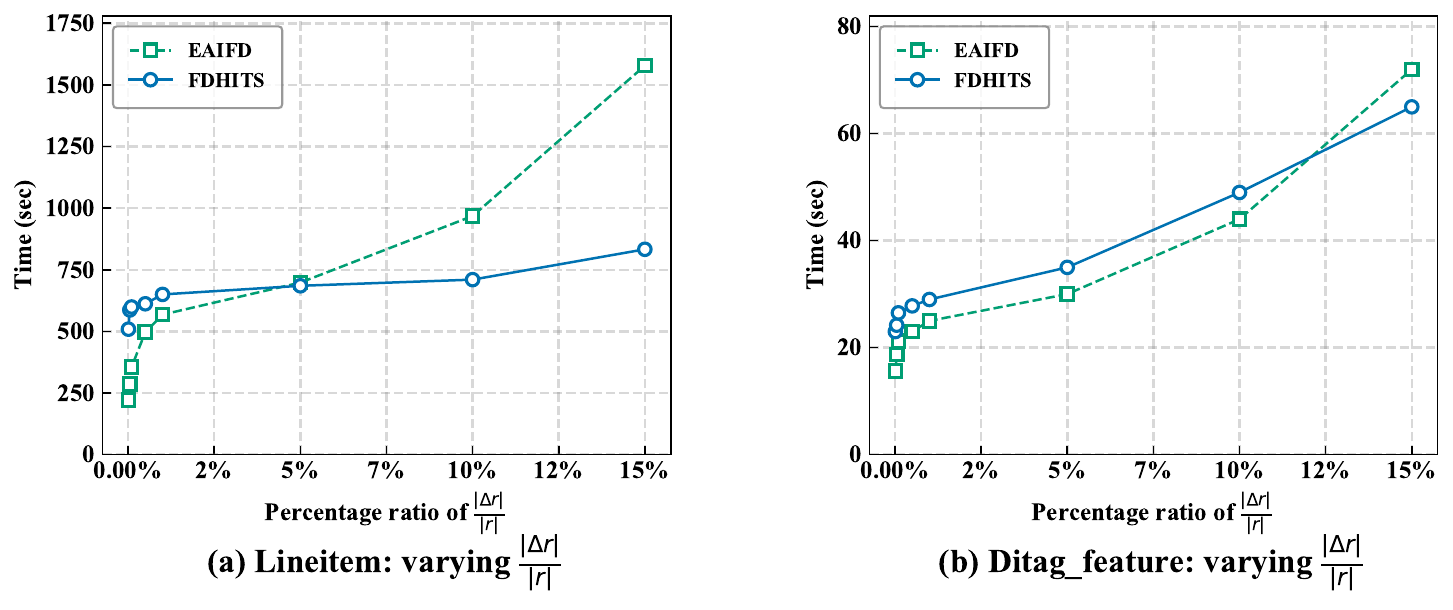}
	\caption{\small Time performance of EAIFD compared with FDHITS across different incremental ratios $\frac{|\Delta r|}{r}$. Subfigures (a) and (b) correspond to Lineitem and Ditag\_feature, respectively.}
	\label{fig:exp5}
\end{figure}

\textbf{Exp $5$: EAIFD against static FD discovery algorithm.}  Experiment $5$ evaluates the performance of EAIFD and FDHITS (a state-of-the-art static algorithm) on Lineitem ($|r|=6,000,000$, $|R|=16$) and Ditag\_feature ($|r|=3,960,124$, $|R|=13$). For FDHITS, \textit{sep} variant is selected over \textit{joint} variant for two reasons. First, the original paper~\cite{DBLP:journals/pacmmod/BleifussPBSN24} demonstrates that \textit{sep} and \textit{joint} variants have highly competitive performance, while the former outperforms the latter on many real-world datasets. Second, the performance of \textit{joint} variant is highly sensitive to sampled data characteristics, while \textit{seq} variant provides more stable and predictable performance. We fix $|r|$ and $|R|$, and vary incremental ratios $\frac{|\Delta r|}{r}$ across a wide range ($0.01\%, 0.05\%, 0.1\%, 0.5\%, 1\%, 5\%, 10\%, 15\%$) to identify a performance break-even point between EAIFD and FDHITS.

The results in Figure \ref{fig:exp5}(a) and (b) show two distinct trends. The runtime of FDHITS is high but grows modestly across all incremental ratios, since its runtime is determined by the total dataset size ($|r| + |\Delta r|$). In contrast, EAIFD shows significant efficiency advantages in small increments (Lineitem: $\frac{|\Delta r|}{r} \le 5\%$; Ditag\_feature: $\frac{|\Delta r|}{r} \le 10\%$), achieving much lower runtime. As the incremental ratio increases, the runtime of EAIFD grows rapidly and eventually exceeds that of FDHITS. This is because the runtime of EAIFD is primarily determined by the incremental data size ($|\Delta r|$). Nevertheless, this experiment confirms the practical value of EAIFD: in typical applications, the incremental update $\Delta r$ typically constitutes a small portion of the large and growing data set $r$. Thus, EAIFD consistently stays on the favorable side of the performance break-even point, making it more suitable for real-world scenarios.

\begin{table}[!htb]
	\centering
	\caption{\small Preprocessing of three algorithms on different datasets.}
	\label{tab:preprocessingTime}
	\renewcommand{\arraystretch}{1.0}
	\setlength{\tabcolsep}{4pt}
	\begin{tabular}{c|c|c|c|c|c|c}
		\hline
		\multicolumn{4}{c|}{\textbf{Dataset Properties}} & \multicolumn{3}{c}{\textbf{Preprocessing Time}} \\ \hline
		\textbf{DataSet} & $\boldsymbol{|r|}$ & $\boldsymbol{|R|}$& $\boldsymbol{|\mathcal{F}|}$ & \textbf{DynFD} & \textbf{DHSFD} & \textbf{EAIFD} \\ 
				
		\hline
		Iris       & 147     & 5    &4   & 0.06s     & 0.03s  & \textbf{0.017s} \\ 
		Bridges   & 108     & 13   &142   & 0.15h     & 1.1s   & \textbf{0.77s}  \\ 
		Hepatitis   & 155     & 20   &8250   & 0.45h     & 1.6s   & \textbf{0.88s}  \\ 
		Horse      & 300     & 29   &128726   & 5.1h     & 2.5s   & \textbf{0.63s}  \\ 
		Balance    & 625     & 5   &1   &\textbf{0.02s}    & 0.62s  & 0.03s  \\ 
		NCV        & 1000    & 19   &758   & 67.4s     & 4.6s   & \textbf{0.7s}   \\ 
		Plista     & 1001    & 63   &173409   & 5.5h     &6.7s   & \textbf{2.8s}   \\ 
		Claim      & 20000   & 11   &12   &128s    &123s    & \textbf{62s}   \\ 
		Letter     & 20000   & 17   &61   & 164s     & 145s   & \textbf{92s}    \\ 
		CAB        & 67300   & 54   &3353531   & 6.5h     & 1.34h  & \textbf{0.77h}  \\ 
		Bioentry   & 184292  & 9   &19   & 0.25h    & 0.42h  & \textbf{0.12h } \\ 
		Flight     & 250000  & 31   &117367   & 6.6h     & 3.45h  & \textbf{0.34h}  \\ 
		Pitches    & 250000  & 40   &608928   & 7.9h     & 5.6h   & \textbf{0.56h}  \\ 
		Ditag\_Feature & 3960124 & 13   &58   &7.2h   & 7.4h    & \textbf{0.67h}   \\ 
		\hline
	\end{tabular}
\end{table}

\textbf{Exp $6$: The preprocessing time of algorithms.} Experiment $6$ evaluates the one-time preprocessing cost of EAIFD, DynFD, and DHSFD on real-world datasets. The startup mechanisms of the three algorithms differ significantly. DynFD computes positive and negative covers from the initial dataset. DHSFD compares all tuple pairs to build the hypergraph. In contrast, EAIFD first sorts the dataset by attributes and performs pairwise comparisons only on sampled data.

Experimental results in Table~\ref{tab:preprocessingTime} show that EAIFD achieves shortest preprocessing times, often over an order of magnitude faster on large datasets. DynFD is slow on large datasets with many FDs, while DHSFD can take several hours on long datasets with many tuples, making startup virtually impractical on very large datasets. EAIFD avoids building auxiliary structures such as PLIs or positive and negative covers. Instead, it sorts each attribute ($O(|R| \times |r| \log|r|)$) and performs pairwise comparisons only on a small sampled dataset ($O(|s|^2)$). The limited sample size keeps preprocessing efficient, enabling rapid initialization in practice.

\textbf{Exp $7$: The memory usage of auxiliary structures.} Experiment $7$ evaluates the memory usage of PLIs and $MHT$ (with $\theta=80\%$) of $r$. As PLIs are the common validation structures, they provide a key benchmark. To analyze the factors influencing memory usage, we record $|r|$, $|R|$, $|\mathcal{F}|$, and memory size of datasets. Results are summarized in Table~\ref{tab:aux_structures}.

Experimental results show that PLI memory usage grows rapidly with increasing $|r|$ and $|R|$, due to maintaining complete attribute partition indices. In contrast, $MHT$ employs a \textit{high-frequency mapping items preservation strategy}, caching only key value mapping items of valid FDs exceeding a frequency threshold $\theta$. Thus each FD stores at most $O(\frac{1}{\theta})$ mapping items, making the overall space complexity $O(\frac{|\mathcal{F}|}{\theta})$, which is independent of dataset size. Consequently, on datasets with few FDs, such as Claim and Bioentry, $MHT$ consumes substantially less memory than PLIs, and its memory usage remains controllable even for large-scale datasets.

On complex datasets or with many FDs, the memory usage of $MHT$ may exceed that of PLIs, which also aligns perfectly with its $O(\frac{|\mathcal{F}|}{\theta})$ space complexity. Nevertheless, this is our time-memory tradeoff design, where EAIFD accepts moderate memory for a substantial performance speedup.

\begin{table}[htbp]
	\centering
	\caption{\small Auxiliary structures on different datasets.}
	\label{tab:aux_structures}
	\renewcommand{\arraystretch}{1.0} % 增加行高
	\setlength{\tabcolsep}{1pt} % 增加列间距
	\begin{tabular}{c|c|c|c|c|c|c}
		\hline
		\multicolumn{4}{c|}{\textbf{Dataset Properties}} & \multicolumn{3}{c}{\textbf{Auxiliary Structures}} \\ 
		\hline
		\textbf{DataSet} & $\boldsymbol{|r|}$ & $\boldsymbol{|R|}$ & $\boldsymbol{|\mathcal{F}|}$ & \textbf{Data} & \textbf{PLIs} & \textbf{MHT ($\boldsymbol{\theta = 80\%}$)} \\ 
		\hline
		Iris & 147 & 5 & 4 & 5KB & 1.7KB & \textbf{0.04KB} \\ 
		Bridges & 108 & 13 & 142 & 6KB & 3KB & \textbf{1.4KB} \\ 
		Hepatitis & 155 & 20 & 8250 & 6.1KB & \textbf{3.6KB} & 8.4KB \\ 
		Horse & 300 & 29 & 128726 & 18KB & \textbf{6.3KB }& 1.2MB \\ 
		Balance & 625 & 5 & 1 & 7KB & 2.3KB & \textbf{0} \\ 
		NCV & 1000 & 19 & 758 & 151KB & 76KB & \textbf{10.3KB} \\ 
		Plista & 1001 & 63 & 173409 & 496KB & \textbf{312KB} & 1.7MB \\ 
		Claim & 20000 & 11 & 12 & 2.6MB & 1.5MB & \textbf{5KB} \\ 
		Letter & 20000 & 17 & 61 & 696KB & 221KB & \textbf{15KB} \\ 
		CAB & 67300 & 54 & 3353531 & 11.4MB & \textbf{8.9MB} & 98MB \\ 
		Bioentry & 184292 & 9 & 19 & 24MB & 16.7MB & \textbf{6KB} \\ 
		Flight & 250000 & 31 & 117367 & 19.7KB & \textbf{11KB} & 1.1MB \\ 
		Pitches & 250000 & 40 & 608928 & 50MB & 27.3MB & \textbf{19MB} \\ 
		Ditag\_Feature & 3960124 & 13 & 58 & 348MB & 202MB & \textbf{14KB} \\ 
		\hline
	\end{tabular}
\end{table}

\textbf{Exp $8$: The effect of frequency threshold $\theta$.} Experiment~8 evaluates the scalability of EAIFD under different mapping frequency thresholds $\theta$ for retaining high-frequency items in the $MHT$. The experiment uses four datasets: Plista, Pitches, CAB, and Flight, with an incremental ratio of $\frac{|\Delta r|}{|r|} = 10\%$.

Table~\ref{tab:differentTheta} illustrates the performance and memory consumption of the $MHT$ for $r$ when the frequency threshold $\theta$ increases from $70\%$ to $90\%$. When $\theta = 70\%$, EAIFD achieves the highest efficiency with the largest $MHT$ memory usage, since more high-frequency mapping items are retained in the initial $MHT$. This allows a large portion of validations to be efficiently performed via hash lookups with $MHT_{\Delta}$, without revalidating on the initial data blocks. This reflects a trade-off between space consumption and time performance. As $\theta$ increases, fewer mapping items satisfy the threshold, leading to reduced memory usage of the $MHT$. Since fewer candidates can be verified by comparing items in the $MHT$, more data blocks need to be loaded and validated, thereby increasing the runtime.

The results clearly demonstrate that $\theta$ is a highly sensitive parameter for the time–memory trade-off. Considering both $MHT$ memory usage and runtime, the results confirm that $\theta = 80\%$ is a well-balanced parameter choice. Overall, the \textit{high-frequency mapping item preservation strategy} enables EAIFD to balance storage efficiency and computational performance.

\begin{table}[t]
	\renewcommand{\arraystretch}{1.2}
	\centering
	\caption{\small Runtime and memory usage of $MHT$ under different mapping frequency $\theta$ on four datasets.}
	\label{tab:differentTheta}
	\resizebox{\columnwidth}{!}{
		\setlength{\tabcolsep}{3pt}
		\begin{tabular}{c|ccccc|ccccc}
			\hline
			\multirow{2}{*}{\textbf{Dataset}} &
			\multicolumn{5}{c|}{\textbf{Runtime (sec)}} &
			\multicolumn{5}{c}{\textbf{Memory of $MHT$ (MB)}} \\
			\cline{2-11}
			& 70\% & 75\% & 80\% & 85\% & 90\% 
			& 70\% & 75\% & 80\% & 85\% & 90\% \\ 
			\hline
			Plista & 0.09 & 0.32 & 0.50 & 2.91 & 4.51 
			& 1.90 & 1.75 & 1.70 & 1.43 & 1.27 \\
			Flight & 42   & 68   & 83   & 174  & 406  
			& 3.27 & 1.93 & 1.10 & 0.82 & 0.56 \\
			Pitches& 72   & 95   & 132  & 357  & 630  
			& 54.7 & 31.6 & 19.0 & 13.6 & 10.8 \\
			CAB    & 78   & 144  & 197  & 522  & 927  
			& 272.6  & 134.2  & 98.7   & 77.3   & 45.5\\
			\hline
	\end{tabular}}
\end{table}

\textbf{Summary of Experimental Analysis.} The experimental results demonstrate that EAIFD achieves superior performance in incremental FD discovery. By reformulating FD discovery as a partial hypergraph hitting set enumeration and employing $MHT$ to preserve historical computations combined with efficient validation methods, EAIFD outperforms existing static and incremental algorithms. It maintains stable performance across varying tuple, attribute, and incremental data scales, while reducing preprocessing time. Moreover, the \textit{high-frequency mapping items preservation strategy} of $MHT$ balances memory and computational efficiency, and the core frequency threshold $\theta$ shows acceptable performance across a wide range. Overall, EAIFD provides an efficient, scalable, and practical solution for incremental FD discovery.

\section{Conclusion}\label{sec:conclusion}

This paper proposes EAIFD, an efficient algorithm for FD discovery in relational databases under continuous incremental updates. EAIFD overcomes redundant cost caused by re-execution in static algorithms and the performance bottlenecks of existing incremental methods. It introduces two core components: (1) modeling incremental FD discovery as hitting set enumeration over partial hypergraphs, (2) employing efficient two-step incremental validation strategy. This design achieves high performance with low memory consumption.

Extensive experiments demonstrate that EAIFD consistently outperforms existing incremental FD discovery (achieving up to an order-of-magnitude speedup on datasets such as Plista) and shows significant performance compared with static algorithms when the incremental data ratio is small. Its efficiency advantage stems from its core operations, which depend primarily on the incremental data size $|\Delta r|$ rather than the full dataset size $|r|$. This characteristic further enhances the efficiency and scalability in long-term applications where $\frac{|\Delta r|}{r}$ decreases over time. In terms of memory, EAIFD achieves a low memory consumption independent of $|r|$ by utilizing $MHT$ with the \textit{high-frequency mapping items preservation strategy}. Furthermore, EAIFD enables fast initialization with preprocessing complexity $O(|R|\times|r|\log|r|)$.

While EAIFD achieves remarkable progress in incremental FD discovery, several directions remain for future work, such as deletions and modifications for fully dynamic FD discovery, and explore parallel or distributed implementations to improve scalability on massive datasets.

% Can use something like this to put references on a page
% by themselves when using endfloat and the captionsoff option.
\ifCLASSOPTIONcaptionsoff
  \newpage
\fi

% trigger a \newpage just before the given reference
% number - used to balance the columns on the last page
% adjust value as needed - may need to be readjusted if
% the document is modified later
%\IEEEtriggeratref{8}
% The "triggered" command can be changed if desired:
%\IEEEtriggercmd{\enlargethispage{-5in}}

% references section

% can use a bibliography generated by BibTeX as a .bbl file
% BibTeX documentation can be easily obtained at:
% http://mirror.ctan.org/biblio/bibtex/contrib/doc/
% The IEEEtran BibTeX style support page is at:
% http://www.michaelshell.org/tex/ieeetran/bibtex/
%\bibliographystyle{IEEEtran}
% argument is your BibTeX string definitions and bibliography database(s)
%\bibliography{IEEEabrv,../bib/paper}
%
% <OR> manually copy in the resultant .bbl file
% set second argument of \begin to the number of references
% (used to reserve space for the reference number labels box)
\bibliographystyle{IEEEtran}
\bibliography{IEEEabrv,reference}

% Generated by IEEEtran.bst, version: 1.14 (2015/08/26)
\begin{thebibliography}{10}
\providecommand{\url}[1]{#1}
\csname url@samestyle\endcsname
\providecommand{\newblock}{\relax}
\providecommand{\bibinfo}[2]{#2}
\providecommand{\BIBentrySTDinterwordspacing}{\spaceskip=0pt\relax}
\providecommand{\BIBentryALTinterwordstretchfactor}{4}
\providecommand{\BIBentryALTinterwordspacing}{\spaceskip=\fontdimen2\font plus
\BIBentryALTinterwordstretchfactor\fontdimen3\font minus
  \fontdimen4\font\relax}
\providecommand{\BIBforeignlanguage}[2]{{%
\expandafter\ifx\csname l@#1\endcsname\relax
\typeout{** WARNING: IEEEtran.bst: No hyphenation pattern has been}%
\typeout{** loaded for the language `#1'. Using the pattern for}%
\typeout{** the default language instead.}%
\else
\language=\csname l@#1\endcsname
\fi
#2}}
\providecommand{\BIBdecl}{\relax}
\BIBdecl

\bibitem{DBLP:books/mg/SKS20}
A.~Silberschatz, H.~F. Korth, and S.~Sudarshan, \emph{Database System Concepts,
  Seventh Edition}.\hskip 1em plus 0.5em minus 0.4em\relax McGraw-Hill Book
  Company, 2020.

\bibitem{DBLP:conf/edbt/PapenbrockN17}
T.~Papenbrock and F.~Naumann, ``Data-driven schema normalization,'' in
  \emph{Proc. 20th Int. Conf. Extending Database Technol.}, 2017, pp. 342--353.

\bibitem{DBLP:journals/tods/WeiL21}
Z.~Wei and S.~Link, ``Embedded functional dependencies and data-completeness
  tailored database design,'' \emph{{ACM} Trans. Database Syst.}, vol.~46,
  no.~2, pp. 7:1--7:46, 2021.

\bibitem{DBLP:journals/vldb/KossmannPN22}
J.~Kossmann, T.~Papenbrock, and F.~Naumann, ``Data dependencies for query
  optimization: a survey,'' \emph{{VLDB} J.}, vol.~31, no.~1, pp. 1--22, 2022.

\bibitem{DBLP:conf/icde/BohannonFGJK07}
P.~Bohannon, W.~Fan, F.~Geerts, X.~Jia, and A.~Kementsietsidis, ``Conditional
  functional dependencies for data cleaning,'' in \emph{Proc. Int. Conf. Data
  Eng.}, 2007, pp. 746--755.

\bibitem{DBLP:books/daglib/0029346}
A.~Doan, A.~Y. Halevy, and Z.~G. Ives, \emph{Principles of Data
  Integration}.\hskip 1em plus 0.5em minus 0.4em\relax Morgan Kaufmann, 2012.

\bibitem{DBLP:journals/cj/HuhtalaKPT99}
Y.~Huhtala, J.~K{\"{a}}rkk{\"{a}}inen, P.~Porkka, and H.~Toivonen, ``{TANE:} an
  efficient algorithm for discovering functional and approximate
  dependencies,'' \emph{Comput. J.}, vol.~42, no.~2, pp. 100--111, 1999.

\bibitem{DBLP:journals/is/NovelliC01}
N.~Novelli and R.~Cicchetti, ``Functional and embedded dependency inference: a
  data mining point of view,'' \emph{Inf. Syst.}, vol.~26, no.~7, pp. 477--506,
  2001.

\bibitem{DBLP:conf/cikm/AbedjanSN14}
Z.~Abedjan, P.~Schulze, and F.~Naumann, ``{DFD:} efficient functional
  dependency discovery,'' in \emph{Proc. {ACM} Int. Conf. Inf. Knowl. Manag.},
  2014, pp. 949--958.

\bibitem{DBLP:conf/sigmod/PapenbrockN16}
T.~Papenbrock and F.~Naumann, ``A hybrid approach to functional dependency
  discovery,'' in \emph{Proc. 2016 {ACM} Int. Conf. Manag. Data}, 2016, pp.
  821--833.

\bibitem{DBLP:journals/pacmmod/BleifussPBSN24}
T.~Bleifu{\ss}, T.~Papenbrock, T.~Bl{\"{a}}sius, M.~Schirneck, and F.~Naumann,
  ``Discovering functional dependencies through hitting set enumeration,''
  \emph{Proc. {ACM} Manag.Data}, vol.~2, no.~1, pp. 43:1--43:24, 2024.

\bibitem{DBLP:journals/dke/LiuYLW13}
J.~Liu, F.~Ye, J.~Li, and J.~Wang, ``On discovery of functional dependencies
  from data,'' \emph{Data Knowl. Eng.}, vol.~86, pp. 146--159, 2013.

\bibitem{DBLP:journals/tkde/WanHWL24}
X.~Wan, X.~Han, J.~Wang, and J.~Li, ``Efficient discovery of functional
  dependencies on massive data,'' \emph{{IEEE} Trans. Knowl Data Eng.},
  vol.~36, no.~1, pp. 107--121, 2024.

\bibitem{DBLP:conf/dawak/WyssGR01}
C.~M. Wyss, C.~Giannella, and E.~L. Robertson, ``Fastfds: {A} heuristic-driven,
  depth-first algorithm for mining functional dependencies from relation
  instances - extended abstract,'' in \emph{Proc. Int. Conf. Data Warehous.
  Knowl. Discov.}, 2001, pp. 101--110.

\bibitem{DBLP:journals/datamine/YaoH08}
H.~Yao and H.~J. Hamilton, ``Mining functional dependencies from data,''
  \emph{Data Min. Knowl. Discov.}, vol.~16, no.~2, pp. 197--219, 2008.

\bibitem{DBLP:conf/icde/WeiL19}
Z.~Wei and S.~Link, ``Discovery and ranking of functional dependencies,'' in
  \emph{Proc. Int. Conf. Data Eng.}, 2019, pp. 1526--1537.

\bibitem{DBLP:conf/edbt/SchirmerP0NHMN19}
P.~Schirmer, T.~Papenbrock, S.~Kruse, F.~Naumann, D.~Hempfing, T.~Mayer, and
  D.~Neusch{\"{a}}fer{-}Rube, ``Dynfd: Functional dependency discovery in
  dynamic datasets,'' in \emph{Proc. Int. Conf. Extending Database Technol.},
  2019, pp. 253--264.

\bibitem{DBLP:conf/icde/XiaoYTMW22}
R.~Xiao, Y.~Yuan, Z.~Tan, S.~Ma, and W.~Wang, ``Dynamic functional dependency
  discovery with dynamic hitting set enumeration,'' in \emph{Proc. Int. Conf.
  Data Eng.}, 2022, pp. 286--298.

\bibitem{DBLP:journals/aicom/FlachS99}
P.~A. Flach and I.~Savnik, ``Database dependency discovery: {A} machine
  learning approach,'' \emph{{AI} Commun.}, vol.~12, no.~3, pp. 139--160, 1999.

\bibitem{DBLP:conf/edbt/LopesPL00}
S.~Lopes, J.~Petit, and L.~Lakhal, ``Efficient discovery of functional
  dependencies and armstrong relations,'' in \emph{Proc. Int. Conf. Extending
  Database Technol.}, ser. Lect. Notes Comput. Sci., vol. 1777, 2000, pp.
  350--364.

\bibitem{DBLP:journals/pvldb/PapenbrockEMNRZ15}
T.~Papenbrock, J.~Ehrlich, J.~Marten, T.~Neubert, J.~Rudolph,
  M.~Sch{\"{o}}nberg, J.~Zwiener, and F.~Naumann, ``Functional dependency
  discovery: An experimental evaluation of seven algorithms,'' \emph{Proc.
  {VLDB} Endow.}, vol.~8, no.~10, pp. 1082--1093, 2015.

\bibitem{DBLP:books/daglib/0020812}
H.~Garcia{-}Molina, J.~D. Ullman, and J.~Widom, \emph{Database systems - the
  complete book {(2.} ed.)}.\hskip 1em plus 0.5em minus 0.4em\relax Pearson
  Education, 2009.

\bibitem{DBLP:journals/dke/MannilaR94}
H.~Mannila and K.~R{\"{a}}ih{\"{a}}, ``Algorithms for inferring functional
  dependencies from relations,'' \emph{Data Knowl. Eng.}, vol.~12, no.~1, pp.
  83--99, 1994.

\bibitem{DBLP:journals/pvldb/BirnickBFNPS20}
J.~Birnick, T.~Bl{\"{a}}sius, T.~Friedrich, F.~Naumann, T.~Papenbrock, and
  M.~Schirneck, ``Hitting set enumeration with partial information for unique
  column combination discovery,'' \emph{Proc. {VLDB} Endow.}, vol.~13, no.~11,
  pp. 2270--2283, 2020.

\bibitem{DBLP:journals/dam/MurakamiU14}
K.~Murakami and T.~Uno, ``Efficient algorithms for dualizing large-scale
  hypergraphs,'' \emph{Discret. Appl. Math.}, vol. 170, pp. 83--94, 2014.

\bibitem{DBLP:conf/sigmod/AgrawalIS93}
R.~Agrawal, T.~Imielinski, and A.~N. Swami, ``Mining association rules between
  sets of items in large databases,'' in \emph{Proc. 1993 {ACM} Int. Conf.
  Manag. Data}, 1993, pp. 207--216.

\bibitem{DBLP:journals/pvldb/Berti-EquilleHN18}
L.~Berti{-}{\'{E}}quille, H.~Harmouch, F.~Naumann, N.~Novelli, and
  S.~Thirumuruganathan, ``Discovery of genuine functional dependencies from
  relational data with missing values,'' \emph{Proc. {VLDB} Endow.}, vol.~11,
  no.~8, pp. 880--892, 2018.

\end{thebibliography}

% Generated by IEEEtran.bst, version: 1.14 (2015/08/26)

% biography section
% 
% If you have an EPS/PDF photo (graphicx package needed) extra braces are
% needed around the contents of the optional argument to biography to prevent
% the LaTeX parser from getting confused when it sees the complicated
% \includegraphics command within an optional argument. (You could create
% your own custom macro containing the \includegraphics command to make things
% simpler here.)
%\begin{IEEEbiography}[{\includegraphics[width=1in,height=1.25in,clip,keepaspectratio]{mshell}}]{Michael Shell}
% or if you just want to reserve a space for a photo:

% biography section
% 
% If you have an EPS/PDF photo (graphicx package needed) extra braces are
% needed around the contents of the optional argument to biography to prevent
% the LaTeX parser from getting confused when it sees the complicated
% \includegraphics command within an optional argument. (You could create
% your own custom macro containing the \includegraphics command to make things
% simpler here.)
%\begin{IEEEbiography}[{\includegraphics[width=1in,height=1.25in,clip,keepaspectratio]{mshell}}]{Michael Shell}
% or if you just want to reserve a space for a photo:

% You can push biographies down or up by placing
% a \vfill before or after them. The appropriate
% use of \vfill depends on what kind of text is
% on the last page and whether or not the columns
% are being equalized.

%\vfill

% Can be used to pull up biographies so that the bottom of the last one
% is flush with the other column.
%\enlargethispage{-5in}

% that's all folks
\end{document}